\documentclass[usenatbib]{mnras}

\usepackage{aas_macros}
\usepackage{graphicx}
\usepackage{latexsym}
\usepackage{amsfonts,amsmath,amssymb}
\usepackage{url}
\usepackage[utf8]{inputenc}
\usepackage[T1]{fontenc}
\usepackage{longtable}
\usepackage{multirow,booktabs}
\usepackage{epsfig}
\usepackage{txfonts}
\usepackage{xtab}
\usepackage{fp}
\usepackage{booktabs}

\usepackage{pdflscape}	

\newcommand{\logns}  {{log$N$-log$S$}}

\newcommand{\num}  {{795}}	

\newcommand{\blg}  {{728}}	

\newcommand{\nes}  {{42}}	

\newcommand{\sw}  {{25}}	

\FPeval\neswsrces{clip(\nes+\sw)}	

\newcommand{\catc}  {{347}}	

\FPeval\newc{round(\num-\catc:0)}	

\newcommand{\alluniq}  {{536}}	

\FPeval\absnew{round(\num-\alluniq:0)}	

\newcommand{\medsepx}  {{1.08}}	

\newcommand{\medsepc}  {{0.27}}	

\newcommand{\allprevc}  {{1436}}	

\newcommand{\rsrc}  {{43}}	

\newcommand{\xcmatch}  {{387}}	

\newcommand{\xcmatchuniq}  {{352}}	

\newcommand{\xmmfov}  {{979}}	

\FPdiv\matchper{\xcmatchuniq}{\num}
\FPeval\matchper{round(\matchper*100:0)}	

\FPdiv\matchperall{\xcmatch}{\num}
\FPeval\matchperall{round(\matchperall*100:0)}	

\FPeval\leftover{round(\num-\xcmatch:0)}	

\FPeval\leftoverxmm{round(\xmmfov-\xcmatchuniq:0)}	

\newcommand{\flatten}  {{$\approx1.3\times10^{37}$}}	

\newcommand{\area}  {{0.6}}	

\newcommand{\fbandsrces}  {{795}}	

\FPeval\fullbandpct{round(\fbandsrces/\num*100:0)}	

\newcommand{\xmmxtramat}  {{895}}	

\FPeval\xmmxtramatpct{round(\xmmxtramat/\xmmfov*100:0)}	

\newcommand{\es}  {{erg s$^{-1}$}}
\newcommand{\esc}  {{erg s$^{-1}$ cm$^{-2}$}}

\newcommand{\HST}{{\em Hubble Space Telescope}}

\newcommand{\Chandra}{{\em Chandra}}
\newcommand{\Chandras}{{\em Chandra's}}

\newcommand{\Einstein}{{\em Einstein}}

\newcommand{\xmmn}{{\em XMM-Newton}}
\newcommand{\rosat}{{\em ROSAT}}

\newcommand{\CFHT}{{\em Canada-France-Hawaii Telescope}}
\newcommand{\CFHTt}{{\em CFHT}}
\newcommand{\nustar}{{\em NuSTAR}}

\newcommand{\CIAO}   {{\em CIAO}}

\newcommand{\CALDB} {{\em CALDB}}

\defcitealias{johnson06-12}{J12}

\begin{document}

\title[Chandra Catalogue of Point Sources in M31]{X-Rays Beware: The Deepest Chandra Catalogue of Point Sources in M31}
\author[N. Vulic, S. C. Gallagher, \& P. Barmby]{N. Vulic$^1$\thanks{nvulic@uwo.ca}, S. C. Gallagher$^1$, \& P. Barmby$^1$  \\
$^1$Department of Physics \& Astronomy, Western University, London, ON, N6A 3K7, Canada}
\date{Received; Accepted}
\maketitle
\label{firstpage}
\pubyear{2016}

\begin{abstract}	

This study represents the most sensitive \Chandra\ X-ray point source catalogue of M31. Using 133 publicly available \Chandra\ ACIS-I/S observations totalling $\sim1$ Ms, we detected \num\ X-ray sources in the bulge, northeast, and southwest fields of M31, covering an area of $\approx$ \area\ deg$^{2}$, to a limiting unabsorbed $0.5-8.0$ keV luminosity of $\sim10^{34}$ \es.
In the inner bulge, where exposure is approximately constant, X-ray fluxes represent average values because they were determined from many observations over a long period of time. Similarly, our catalogue is more complete in the bulge fields since monitoring allowed more transient sources to be detected.
The catalogue was cross-correlated with a previous \xmmn\ catalogue of M31's $D_{25}$ isophote consisting of 1948 X-ray sources, with only 979 within the field of view of our survey. We found \xcmatch\ (\matchperall\%) of our \Chandra\ sources (\xcmatchuniq\ or \matchper\% unique sources) matched to within 5\arcsec\ of \xcmatchuniq\ \xmmn\ sources. Combining this result with matching done to previous \Chandra\ X-ray sources we detected \absnew\ new sources in our catalogue. We created X-ray luminosity functions (XLFs) in the soft ($0.5-2.0$ keV) and hard ($2.0-8.0$ keV) bands that are the most sensitive for any large galaxy based on our detection limits. 
Completeness-corrected XLFs show a break around \flatten\ \es, consistent with previous work. As in past surveys, we find the bulge XLFs are flatter than the disk, indicating a lack of bright high-mass X-ray binaries in the disk and an aging population of low-mass X-ray binaries in the bulge.

\end{abstract}
\begin{keywords}
catalogues --- galaxies: individual: M31, NGC 224 --- X-rays: binaries --- X-rays: galaxies
\end{keywords}

\section{Introduction} \label{sec:intro}

M31 is the nearest large spiral galaxy to our own, and as such allows for the best possible spatial resolution and sensitivity of any Milky Way sized galaxy. The X-ray population of spiral galaxies can include X-ray binaries (XRBs), both low-mass (LMXBs) and high-mass (HMXBs), supernova remnants, supersoft sources, and massive stars. There is also contamination from background active galactic nuclei (AGN) or galaxies/galaxy clusters and Galactic foreground stars. Unlike the Milky Way, which is difficult to observe in X-rays due to absorption and the fact we are in the disc, M31 has moderate Galactic foreground absorption ($N_{\rm{H}} = 7\times10^{20}$ cm$^{-2}$ \citealt{dickey-90}) and can provide a galaxy-wide survey of the X-ray population, specifically XRBs. With an increased sample size, it is possible to study large-scale properties like the radial distribution across a range of $L_{X}$. The goal of this paper is to create the deepest \Chandra\ X-ray catalogue of regions with archival observations in M31. Specifically, the bulge has $>1$ Ms of data largely from monitoring programs that have not been fully exploited.

The X-ray point source population of M31 was first studied by \citep{trinchieri11-91} using $\approx$ 300 ks of \Einstein\ imaging observations. They detected 108 sources of which 16 showed variability. They did not find a significant difference between the luminosity distribution of the bulge and disc population. \citet{primini06-93} completed a survey of the central $\sim$34\arcmin\ of M31 using the \rosat\ High-Resolution Imager (HRI). They found 18 variable sources within $7.5\arcmin$ of the nucleus and 3 probable transients. Also, $>75\%$ of the unresolved X-ray emission in the bulge was either thought to be diffuse of from a new class of X-ray sources. This work was followed-up by deeper \rosat\ observations \citep{supper01-97,supper07-01} detecting 560 sources down to $5\times10^{35}$ \es\ in the $\sim10.7$ deg$^{2}$ view. They associated 55 sources with foreground stars, 33 with globular clusters, 16 with supernova remnants, and 10 with radio sources and galaxies, leaving 80\% of sources without an optical/radio identification. They confirmed the previous result from \Einstein\ that the total luminosity is distributed evenly between the bulge and disc. A comparison with the \Einstein\ results revealed 11 variable sources and 7 transients, while comparison with the first \rosat\ Position Sensitive Proportional Counter (PSPC) survey found 34 variable sources and 8 transients. The \rosat\ surveys also revealed the presence of supersoft sources (a class of white dwarf X-ray binary) in M31 \citep{supper01-97,kahabka04-99}. \citet{trudolyubov12-04} used \xmmn\ and \Chandra\ to detect 43 X-ray sources coincident with globular cluster candidates, finding 31 of the brightest had spectral properties similar to Galactic LMXBs. X-ray monitoring of optical novae in the centre of M31 with \rosat, \xmmn, and \Chandra\ \citep{pietsch11-05,pietsch04-07} showed them to be primarily supersoft sources. \citet{shaw-greening03-09} completed an \xmmn\ spectral survey of 5 fields along the major axis of M31 (excluding the bulge) and detected 335 X-ray sources, which were correlated with earlier X-ray surveys and radio, optical, and infrared catalogues. They classified 18 sources as HMXB candidates by spectral fitting with a power law model with a photon index of $0.8-1.2$, indicating they were magnetically accreting neutron stars. \citet{peacock10-10} used the 2XMMi catalogue of X-ray point sources along with supplemental \Chandra\ and \rosat\ observations to identify 45 globular cluster LMXBs. This study covered 80\% of the known globular clusters in M31 \citep{peacock07-10} and confirmed trends whereby high metallicity, luminosity, and stellar collision rate correlated positively with the likelihood of a cluster hosting an LMXB. \citet{henze03-14} completed monitoring observations with \xmmn\ and \Chandra\ of the bulge of M31 and detected 17 new X-ray counterparts of optical novae, with 24 detected in total.

\begin{table*}
\caption{Summary of Previous M31 X-ray Surveys\label{tab:sum}}
\resizebox{\textwidth}{!}{%

\begin{tabular}{ c c c c c}
\hline\hline
Observatory	&	Detected Sources	&	$L_{X}$ (\es)	&	Region	&	References	\\
\hline
\Einstein	&	108	&	$5\times10^{36} - 10^{38}$ ($0.2-4.0$ keV)	&	14 \Einstein\ imaging observations ($\sim4$ deg$^2$)	&	\citet{trinchieri11-91}	\\
\rosat\ (HRI)	&	86	&	$\gtrsim1.8\times10^{36}$ ($0.2-4.0$ keV)	&	central $\sim$34\arcmin\ ($\sim0.3$ deg$^2$)	&	\citet{primini06-93}\\
\rosat\ (PSPC)	&	560	&	$5\times10^{35} - 5.5\times10^{38}$ ($0.1-2.4$ keV)	&	whole galaxy ($>D_{25}$ ellipse, $10.7$ deg$^2$)	&	\citet{supper01-97,supper07-01}\\
\xmmn/\Chandra	&	43		&	$\sim10^{35} - 10^{39}$ ($0.3-10.0$ keV)	&	bulge \& major axis ($1.7$ deg$^{2}$)	&	\citet{trudolyubov12-04}\\	

\xmmn	&	335		&	$\sim10^{34} - 10^{39}$ ($0.3-10.0$ keV)	&	5 fields along major axis ($1$ deg$^{2}$)	&	\citet{shaw-greening03-09}\\
\xmmn/\Chandra	&	45		&	$\sim10^{35} - 7\times10^{38}$ ($0.2-12.0$ keV)	&	whole galaxy ($>D_{25}$ ellipse, $4$ deg$^{2}$)	&	\citet{peacock10-10}\\
\xmmn	&	1897$^{1}$		&	$4.4\times10^{34} - 2.7\times10^{38}$ ($0.2-4.5$ keV)	&	whole galaxy ($>D_{25}$ ellipse, $4$ deg$^{2}$)	&	\citet{stiele10-11}\\
\xmmn/\Chandra\	&	24		&	$\sim10^{35} - 9\times10^{37}$ ($0.2-2.0$ keV)	&	centre ($\sim0.2$ deg$^{2}$)	&	\citet{henze03-14}\\	

\hline
\end{tabular}
}
\begin{list}{}{}
\item Luminosities have all been corrected to a distance of 776 kpc used in this paper. \Chandra\ surveys are summarized in Table \ref{tab:chandra}. See \citet{stiele10-11} for a more comprehensive list.
\item ${^1}$ The \xmmn\ LP total catalogue includes 1948 X-ray sources.
\end{list}
\end{table*}

The most comprehensive X-ray population survey of M31 to date was completed by \citet{stiele10-11} using the \xmmn\ European Photon Imaging Camera. They detected 1897 sources to a limiting luminosity of $4.4\times10^{34}$ \es, including 914 new X-ray sources. Their source classification/identification was based on several methods: X-ray hardness ratios, spatial extent of the sources, long-term X-ray variability, and cross-correlation with X-ray, optical, infrared, and radio catalogues. Confirmed identifications included 25 supernova remnants, 46 LMXBs, 40 foreground stars, and 15 AGN/galaxies. There were many candidates for each of these classes as well, including 2 HMXBs and 30 supersoft sources. Nevertheless, 65\% of their sources had no classification. We summarize a few of the major X-ray surveys of M31 in Table \ref{tab:sum} (for a more detailed list please see \citet{stiele10-11}). 

\Chandra\ has not observed all of M31 as previous observatories have, but instead mostly monitored the supermassive black hole in the nucleus, with the majority of exposures each being 5 ks. Various groups have used a handful of observations to survey the bulge and create a catalogue of sources with either the Advanced CCD Imaging Spectrometer (ACIS-I/S) or the High-Resolution Camera (HRC-I/S). \citet{kong10-02} compiled the first \Chandra\ catalogue of M31 within the bulge, finding 204 sources above $\gtrsim2\times10^{35}$ \es. Their most important result was finding different X-ray luminosity functions (XLFs) when different regions (inner/outer bulge and disc) were considered separately. The inner bulge showed a break at $10^{36}$ \es\ and this break shifted to higher luminosities when moving outwards from the inner bulge to the disc. In addition, the slopes became steeper, indicating non-uniform star formation history. \citet{kaaret10-02} used the HRC-I to detect 142 sources, which when compared to \rosat\ observations revealed 50\% of the sources $>5\times10^{36}$ \es\ to be variable. No evidence was found for X-ray pulsars, leading to the conclusion that most sources should be LMXBs. \citet{di-stefano05-02} used 3 disc fields in M31 to analyse globular cluster LMXBs while \citet{di-stefano07-042} used these fields with a nuclear pointing to study supersoft/quasi-soft sources. \citet{williams07-04} used HRC-I to study the disc and bulge of M31 with snapshot images, finding variability in 25\% of 166 detected sources. \citet{voss06-07} combined 26 ACIS observations to investigate the X-ray population in the bulge, finding 263 X-ray sources (64 new) down to $10^{35}$ \es. They clearly demonstrated the power of merging observations to obtain deeper exposures and detect the faintest sources, decreasing the (completeness-corrected) XLF limit by a factor of 3. \citet{hofmann07-13} used 64 HRC-I observations totalling 1 Ms to detect 318 X-ray sources. They studied the long-term variability of sources by producing light curves and found 28 new sources, along with classifying 115 as candidate XRBs. A further 14 globular cluster XRB candidates, several new nova candidates, and a new supersoft X-ray source outburst were discovered. Recently, the $\sim$12 yrs of monitoring observations of the nucleus have been utilized in a number of studies to investigate variability and detect transients \citep{barnard09-12,barnard092-12,barnard06-13}. Specifically, \citet{barnard01-14} used 174 \Chandra\ ACIS and HRC observations to detect 528 X-ray sources in the bulge down to $10^{35}$ \es. By studying source variability, they identified 250 XRBs (200 new) with X-ray data alone, a factor of 4 increase. Table \ref{tab:chandra} summarizes previous \Chandra\ M31 X-ray catalogues. At the time of writing, a large \Chandra\ program (350 ks) has been accepted to survey a part of the star-forming disc of M31 (PI: B. Williams). Aside from being able to confirm the first HMXBs in M31, it will completely characterize a large part of the X-ray source population using optical photometry and spectroscopy.  

\begin{table*}
\caption{Summary of Previous \Chandra\ M31 X-ray Catalogues\label{tab:chandra}}
\begin{tabular}{c c c c c}
\hline\hline
Instrument	&	Detected Sources	&	$L_{X}$ (\es)	&	Region	&	References	\\
\hline
ACIS	-I	&	204	&	$\gtrsim2\times10^{35}$ ($0.3-7.0$ keV)		&	central $\sim17\arcmin\times17\arcmin$ (0.08 deg$^{2}$)	&	\citet{kong10-02}	\\
HRC-I		&	142	&	$2\times10^{35} - 2\times10^{38}$ ($0.1-10.0$ keV)		&	central $\sim30\arcmin\times30\arcmin$ (0.25 deg$^{2}$)	&	\citet{kaaret10-02}\\
ACIS	-I/S \& HRC-I	&	28	&	$5\times10^{35} - 3\times10^{38}$ ($0.3-7.0$ keV)	&	3 disc fields (0.7 deg$^{2}$)	&	\citet{di-stefano05-02}\\
ACIS	-S3	&	33	&	$\sim10^{35} - 10^{38}$ ($0.1-7.0$ keV)	&	3 disc fields and nucleus	(0.7 deg$^{2}$) &	\citet{di-stefano07-042}\\
HRC-I	&	166	&	$\sim10^{36} - 5\times10^{38}$ ($0.1-10.0$ keV)	&	disc \& bulge (0.9 deg$^{2}$)	&	\citet{williams07-04}\\
ACIS	-I/S	&	263	&	$5\times10^{33} - 2\times10^{38}$ ($0.5-8.0$ keV)	&	$12$\arcmin\ radius from core (0.126 deg$^{2}$)	&	\citet{voss06-07}\\
HRC-I	&	318	&	N/A ($0.1-10.0$ keV)	&	$30$\arcmin\ radius from core ($0.8$ deg$^{2}$) &	\citet{hofmann07-13}\\
ACIS	-I/S \& HRC-I	&	528	&	$5\times10^{34} - 5\times10^{38}$ ($0.3-10.0$ keV)	&	$20$\arcmin\ radius from core ($0.35$ deg$^{2}$) &	\citet{barnard01-14}\\
\hline
\end{tabular}
\begin{list}{}{}
\item Luminosities have all been corrected to a distance of 776 kpc used in this paper.
\end{list}
\end{table*}

This paper aims to study the properties of M31 X-ray point sources down to the lowest luminosities for any large galaxy. Specifically, we will report general source catalogue characteristics (e.g. flux and radial distributions), cross-correlate our catalogue with previous \xmmn\ and \Chandra\ surveys, and study the XLF. In addition, high-resolution \HST\ data from the Panchromatic Hubble Andromeda Treasury (PHAT) Survey \citep{dalcanton06-12} allows optical counterpart identifications for X-ray sources, especially AGN. We adopt a distance to M31 of 776 $\pm$ 18 kpc as in \citet{dalcanton06-12}, which corresponds to a linear scale of 3.8 pc arcsecond$^{-1}$.

\section{Observations and Data Reduction} \label{sec:obs}

\subsection{Observations and Preliminary Reduction} \label{sec:datared1}

We used all 133 publicly available \Chandra\ ACIS observations of M31, composed of 29 ACIS-S and 104 ACIS-I observations, summarized in Tables \ref{tab:acis-s} and \ref{tab:acis-i} respectively. We only used observations where M31 was the target, and so overlapping observations where M32 was the target (e.g.\ ObsIDs 2017, 2494, 5690) were not included. \citet{revnivtsev10-07} published an X-ray point source catalogue of M32 by merging all available \Chandra\ ACIS observations in the field.
Data reduction was performed using the \Chandra~Interactive Analysis of Observations (\CIAO) tools package version 4.5 \citep{fruscione07-06} and the \Chandra~Calibration database (\CALDB) version 4.5.5 \citep{graessle07-06}. Before astrometric alignment could be completed we had to create source lists for each of the 133 ACIS observations in order to match them to a reference list. Starting from the level-1 events file, we created a bad pixel file using $\texttt{acis\_run\textunderscore hotpix}$ and eliminated cosmic ray afterglows with only a few events using $\texttt{acis\_detect\_afterglow}$. We then processed the level-1 events file with $\texttt{acis\_process\_events}$ using the default parameters to update charge transfer inefficiency (CTI), time-dependent gain, and pulse height. The VFAINT option was set for certain observations as appropriate. We removed the pixel randomization option by setting $\texttt{pix\_adj}$ to ``none'' to maintain the native position of each photon.
Each  events file was then filtered using the standard (ASCA) grades ($0, 2-4, 6$), status bits ($0$), good time intervals, and CCD chips (I0-I3 for ACIS-I and S3 for ACIS-S). We only used data from the S3 chip from ACIS-S observations due to the degradation of the point spread function (PSF) at large off-axis angles. We then created exposure maps and exposure-corrected images using $\texttt{fluximage}$ with a binsize of 1 in the soft ($0.3-2$ keV), hard ($2-8$ keV), and full ($0.3-8$ keV) energy bands. Source lists in each energy band were created with $\texttt{wavdetect}$ using the $\sqrt{2}$ series from 1 to 8 for the $\textit{scales}$ parameter and corresponding exposure maps to reduce false positives. Default settings were used for all other parameters. We did not remove background flares from our level-2 events files because the background for point sources is addressed in the sensitivity calculation (Section \ref{sec:sens}). We checked for background flares in all level-2 event files (ACIS-I and ACIS-S) using the \CIAO\ $\texttt{deflare}$ script and found negligible flaring throughout. 

\begin{figure*}
\includegraphics[width=1.0\textwidth]{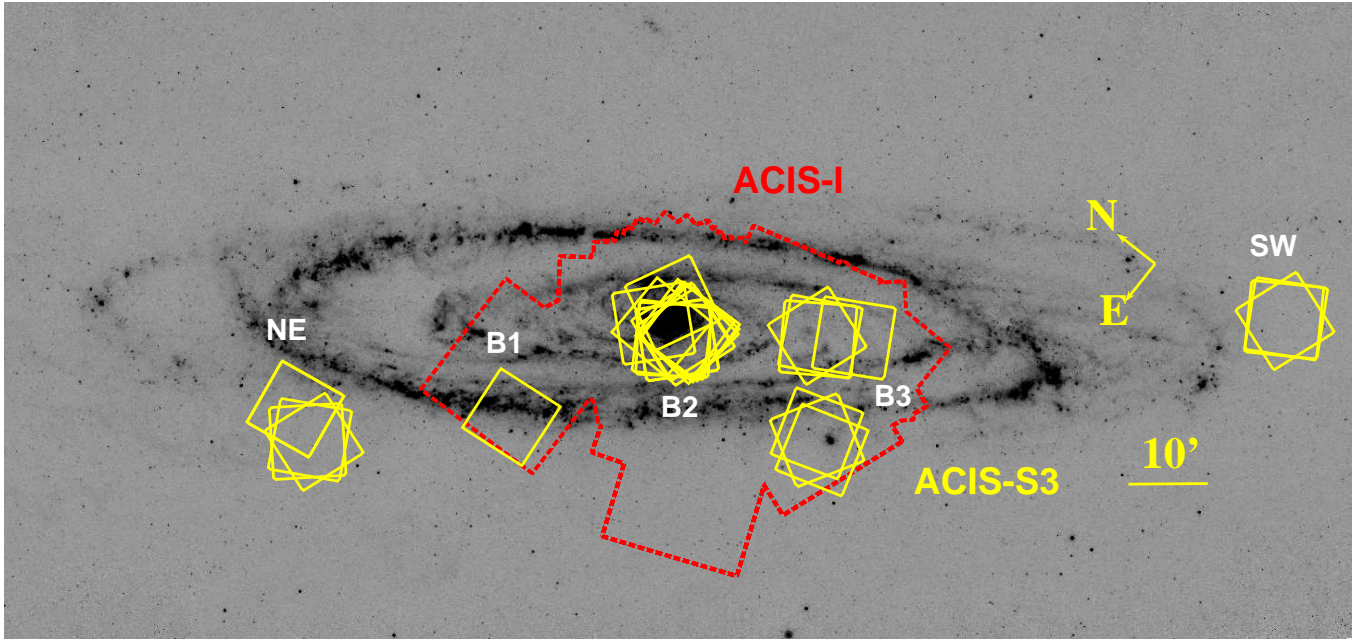}
\caption{The field of view of the ACIS-I (red dashed outline) merged observations and ACIS-S3 chips (yellow) for each observation used in our analysis overlaid on a Spitzer 24 $\micron$ image of M31 \citep{gordon02-06}. Far left is the northeast region (NE), the bulge region is approximated by the red dashed outline (includes the ACIS-S observations within it), and the southwest region is far right (SW). The bulge region is additionally labelled into 3 separate regions based on the distribution of ACIS-S observations, bulge 1 (B1), bulge 2 (B2), and bulge 3 (B3). B3 includes the observations overlapping M32 as well as the fields just above it.} \label{fig:acis-fov}
\end{figure*}

\begin{table*}
\caption{ACIS-S Observations\label{tab:acis-s}}
\resizebox{\textwidth}{!}{%

\begin{tabular}{ccrccccccccccccc}
\hline\hline
ObsID	&	Date	&	R.A.	&	Decl.	&	Distance	&	Livetime	&	Datamode	&	CCDs	&	Roll Angle	&	Region	&	$\delta x$	&	$\delta y$	&	Rotation	&	Scale Factor	&	Matched Sources	&	Catalogue	\\
& & \multicolumn{2}{c}{(J2000)}	&	($\arcmin$) 	& (ks)	& & & ($\degr$)	& &		(Pixels)	&	(Pixels)	&	($\degr$)	& &	&	\\
(1)	&	(2)	&	(3)	&	(4)	&	(5)	&	(6)	&	(7)	&	(8)	&	(9)	&	(10)	&	(11)	&	(12)	&	(13)	&	(14)	&	(15)	&	(16)	\\
\hline
309 & 2000-06-01 & 10.68800 & 41.27877 & 0.6 & 5110 & FAINT & 235678 & 87.5 & B2 & -0.16364 & -0.26018 & -0.03257 & 0.99909 & 4 & CFHT \\
310 & 2000-07-02 & 10.68311 & 41.27876 & 0.6 & 5263 & FAINT & 235678 & 108.7 & B2 & 0.36313 & -0.19187 & 0.10652 & 1.00187 & 8 & CFHT \\
313 & 2000-09-21 & 10.65816 & 40.87837 & 23.5 & 5980 & FAINT & 235678 & 162.9 & B3 & 0.48271 & -0.96208 & 0.03642 & 1.00353 & 3 & 2MASS \\
314 & 2000-10-21 & 10.64900 & 40.86757 & 24.1 & 5190 & FAINT & 235678 & 209.4 & B3 & 0.49169 & -1.32084 & 0.05563 & 1.00032 & 6 & X-ray \\
2052 & 2000-11-01 & 11.53802 & 41.65947 & 45.0 & 13878 & FAINT & 235678 & 226.9 & NE & -0.00785 & -0.60853 & 0.08427 & 1.00164 & 4 & CFHT \\
2046 & 2000-11-05 & 9.62713 & 40.26241 & 77.2 & 14767 & FAINT & 235678 & 237.6 & SW & 1.25572 & -0.05402 & -0.10691 & 0.99895 & 8 & X-ray \\
2049 & 2000-11-05 & 10.44879 & 40.98187 & 20.2 & 14572 & FAINT & 235678 & 236.1 & B3 & -0.42443 & 0.56398 & -0.23159 & 1.00028 & 5 & 2MASS \\
1580 & 2000-11-17 & 10.66379 & 40.85630 & 24.8 & 5407 & FAINT & 235678 & 251.5 & B3 & 0.42793 & -0.92874 & 0.19477 & 1.00102 & 4 & 2MASS \\
1854 & 2001-01-13 & 10.67349 & 41.25548 & 0.9 & 4697 & FAINT & 235678 & 295.6 & B2 & -0.65829 & -0.27766 & 0.09229 & 1.00050 & 5 & CFHT \\
2047 & 2001-03-06 & 9.69689 & 40.27309 & 74.7 & 14591 & FAINT & 235678 & 330.4 & SW & 2.11548 & -1.17785 & -0.01302 & 0.99931 & 7 & X-ray \\
2053 & 2001-03-08 & 11.61538 & 41.66618 & 48.2 & 13363 & FAINT & 235678 & 330.5 & NE & 0.25278 & -1.17593 & -0.09831 & 0.99927 & 4 & CFHT \\
2050 & 2001-03-08 & 10.51841 & 40.99440 & 18.1 & 13046 & FAINT & 235678 & 331.7 & B3 & -1.22735 & -0.54781 & 0.06256 & 1.00063 & 4 & 2MASS \\
2048 & 2001-07-03 & 9.64190 & 40.33115 & 73.5 & 13779 & FAINT & 235678 & 109.6 & SW & -0.25110 & -0.31239 & -0.10373 & 0.99851 & 4 & Nomad \\
2051 & 2001-07-03 & 10.45902 & 41.05868 & 16.2 & 13583 & FAINT & 235678 & 109.0 & B3 & 0.50294 & -0.03155 & -0.18985 & 0.99596 & 5 & 2MASS \\
2054 & 2001-07-03 & 11.54972 & 41.71649 & 47.3 & 14531 & FAINT & 235678 & 108.3 & NE & -0.59834 & -0.00160 & -0.19307 & 0.99860 & 4 & CFHT \\
1575 & 2001-10-05 & 10.66840 & 41.27894 & 1.0 & 39093 & FAINT & 235678 & 180.4 & B2 & 0.21776 & -0.65811 & 0.06315 & 1.00038 & 10 & CFHT \\
2900 & 2002-11-29 & 11.16410 & 41.35916 & 22.3 & 5331 & FAINT & 136789 & 263.7 & B1 & 0.00674 & -1.04238 & -0.15971 & 1.00138 & 4 & SDSS \\
4541 & 2004-11-25 & 11.51934 & 41.70871 & 45.9 & 25631 & FAINT & 235678 & 260.0 & NE & 0.00551 & -0.98102 & -0.00539 & 0.99743 & 7 & SDSS \\
6167 & 2004-11-26 & 11.51931 & 41.70871 & 45.9 & 24230 & FAINT & 235678 & 260.0 & NE & 0.76975 & -0.25521 & -0.04166 & 1.00155 & 7 & SDSS \\
4536 & 2005-03-07 & 10.46483 & 40.93912 & 22.1 & 56181 & FAINT & 235678 & 330.4 & B3 & 0.24710 & -2.32722 & -0.09497 & 1.00424 & 3 & X-ray \\
14197 & 2011-09-01 & 10.68124 & 41.26788 & 0.2 & 37465 & FAINT & 235678 & 143.6 & B2 & -0.18277 & -0.08271 & 0.01628 & 0.99981 & 7 & CFHT \\
14198 & 2011-09-06 & 10.68135 & 41.26769 & 0.2 & 39875 & FAINT & 235678 & 147.7 & B2 & -0.22336 & 0.35097 & -0.00697 & 1.00107 & 9 & CFHT \\
13825 & 2012-06-01 & 10.68170 & 41.27070 & 0.2 & 39791 & FAINT & 23567 & 87.7 & B2 & -0.09198 & 0.00991 & 0.03199 & 1.00160 & 8 & CFHT \\
13826 & 2012-06-06 & 10.68155 & 41.27051 & 0.2 & 36681 & FAINT & 235678 & 91.9 & B2 & -0.32372 & -0.29423 & 0.03545 & 1.00100 & 9 & CFHT \\
13827 & 2012-06-12 & 10.68139 & 41.27031 & 0.2 & 41481 & FAINT & 235678 & 96.2 & B2 & 0.12455 & 1.22787 & -0.10924 & 1.00182 & 8 & CFHT \\
13828 & 2012-07-01 & 10.68111 & 41.26974 & 0.2 & 39881 & FAINT & 235678 & 107.9 & B2 & -0.06387 & -0.33627 & 0.02094 & 1.00162 & 8 & CFHT \\
14195 & 2012-08-14 & 10.68103 & 41.26846 & 0.2 & 27891 & FAINT & 235678 & 132.2 & B2 & 0.49410 & 0.38553 & -0.00897 & 1.00107 & 10 & CFHT \\
15267 & 2012-08-16 & 10.68105 & 41.26846 & 0.2 & 11103 & FAINT & 235678 & 132.2 & B2 & 0.41818 & -0.28493 & 0.01966 & 0.99979 & 8 & CFHT \\
14196 & 2012-10-28 & 10.68565 & 41.26598 & 0.2 & 42851 & FAINT & 23567 & 221.7 & B2 & -0.02078 & -0.09118 & 0.05369 & 1.00091 & 10 & CFHT \\
\hline
\end{tabular}
}
\begin{list}{}{}
\item The R.A. (column 3) and Decl. (column 4) represent the pointing of the optical axis, the RA\_PNT and DEC\_PNT keywords from the level-2 events files. Distance (column 5) represents the distance in arcminutes of the observation's pointing (RA\_PNT and DEC\_PNT) from the centre of M31 (J004244.33+411607.50). Livetime (column 6) is the total amount of time that CCDs actually observe a source, which excludes the dead time (e.g.\ time it takes to transfer charge from the image region to the frame store region). CCDs (column 8) indicates the CCD chips that were on during the observation, where for ACIS-S we only used chip 7 (S3) for analysis. Region (column 10) defines one of the 5 ACIS-S regions in Figure \ref{fig:acis-fov}: northeast (NE), left (northeast) of the bulge (B1), bulge centre (B2), right (southwest) of the bulge (B3), and southwest (SW). Columns $11-16$ summarize the results from matching each ObsID to a reference astrometric catalogue (see Section \ref{sec:imreg} for details) using the \CIAO\ tool $\texttt{wcs\_match}$.
\end{list}
\end{table*}

\begin{table*}
\caption{ACIS-I Observations\label{tab:acis-i}}
\resizebox{\textwidth}{!}{%

\begin{tabular}{ccrcccccccccccc}
\hline\hline
ObsID	&	Date	&	R.A.	&	Decl.	&	Distance	&	Livetime	&	Datamode	&	CCDs	&	Roll Angle	&	$\delta x$	&	$\delta y$	&	Rotation	&	Scale Factor	&	Matched Sources	&	Catalogue	\\
& & \multicolumn{2}{c}{(J2000)}	&	($\arcmin$)	& (ks)	& & & ($\degr$) & (Pixels)	&	(Pixels)	&	($\degr$)	& &	&	\\	
(1)	&	(2)	&	(3)	&	(4)	&	(5)	&	(6)	&	(7)	&	(8)	&	(9)	&	(10)	&	(11)	&	(12)	&	(13)	&	(14)	&	(15)	\\
\hline
303 & 1999-10-13 & 10.67806 & 41.26989 & 0.3 & 11839 & FAINT & 012367 & 193.0 & -0.18797 & 0.16872 & 0.02523 & 0.99995 & 15 & CFHT \\
305 & 1999-12-11 & 10.68273 & 41.26395 & 0.3 & 4131 & FAINT & 01236 & 274.0 & -0.26412 & 0.36347 & 0.02543 & 0.99911 & 12 & CFHT \\
306 & 1999-12-27 & 10.68409 & 41.26370 & 0.3 & 4134 & FAINT & 01236 & 285.4 & -0.00194 & -0.34534 & 0.03549 & 1.00060 & 12 & CFHT \\
307 & 2000-01-29 & 10.68116 & 41.25783 & 0.7 & 4118 & FAINT & 01236 & 304.3 & -0.14524 & -0.12621 & 0.02251 & 0.99985 & 15 & CFHT \\
308 & 2000-02-16 & 10.68763 & 41.25244 & 1.0 & 4015 & FAINT & 012356 & 315.0 & -0.02836 & -0.76721 & -0.03976 & 1.00036 & 13 & CFHT \\
\hline
\end{tabular}
}
\begin{list}{}{}
\item The R.A. (column 3) and Decl. (column 4) represent the pointing of the optical axis, the RA\_PNT and DEC\_PNT keywords from the level-2 events files. Distance (column 5) represents the distance in arcminutes of the observation's pointing (RA\_PNT and DEC\_PNT) from the centre of M31 (J004244.33+411607.50). Livetime (column 6) is the total amount of time that CCDs actually observe a source, which excludes the dead time (e.g.\ time it takes to transfer charge from the image region to the frame store region). CCDs (column 8) indicates the CCD chips that were on during the observation, where for ACIS-I we used chips 0123 (I0-I3) for analysis. Also, ObsID's 1581/82 only had 2 active CCDs. Columns $10-15$ summarize the results from matching each ObsID to a reference astrometric catalogue (see Section \ref{sec:imreg} for details) using the \CIAO\ tool $\texttt{wcs\_match}$. We do not classify ACIS-I observations by region since they are contiguous. 
(This table is published in its entirety in the electronic edition of the journal. A portion is shown here for guidance regarding its form and content.)
\end{list}
\end{table*}

\subsection{Image Registration} \label{sec:imreg}

While \Chandra\ has absolute astrometry of $\approx0.6\arcsec$ (90 per cent uncertainty circle, within 3\arcmin\ of the aimpoint)\footnote{\url{http://cxc.harvard.edu/cal/ASPECT/celmon/}},  creating an X-ray point source catalogue requires more precision. Therefore we performed alignment to a ground-based standard for every ACIS observation to improve astrometry to the extent possible. We began by using a wide-field image that covers the PHAT-fields \citep{williams11-14} obtained with the \CFHT~(\CFHTt) MegaPrime/MegaCam in the $i$ (MP9701) filter. The astrometry in this image had already been corrected to match the Two Micron All Sky Survey (2MASS, \citealt{skrutskie02-06}) reference system, which is accurate to $<0.2\arcsec$. We used this image because the PHAT data was aligned to it, therefore making optical counterpart identification more precise. We used the \texttt{daofind} tool in the \texttt{IRAF\textbackslash DAOPHOT} package \citep{stetson03-87} to compile a source list from the \CFHTt~image. We chose sources in the \CFHTt\ image that were either background galaxies or globular cluster LMXB sources. For the $\texttt{datapars}$ settings we changed the $\it{fwhmpsf}$ parameter to 2 based on an analysis of numerous point sources in the image. The $\it{sigma}$ (the standard deviation of the mode of the background), $\it{datamin}$, and $\it{datamax}$ parameters were all calculated from the image; the $\it{readnoise}$ was set to 5. For $\texttt{centerpars}$ the $\it{calgorithm}$ was set to centroid, while the $\texttt{fitskypars}$ parameter $\it{salgorithm}$ was set to median and $\it{annulus}$ and $\it{dannulus}$ were both set to 8 ($4\times$ $\it{fwhmpsf}$). The $\texttt{photpars}$ parameter $\it{zmag}$ was set to 25.72, the zero point of the magnitude scale for the $i$ filter on the \CFHTt~MegaPrime/MegaCam\footnote{\url{http://www.cfht.hawaii.edu/Instruments/Imaging/MegaPrime/generalinformation.html}}. The $\texttt{findpars}$ settings were left to their defaults. We ran the $\texttt{phot}$ procedure with the above settings to determine precise centroids for our output source list.

We used the \CIAO\ tool $\texttt{wcs\_match}$ to match and then align the X-ray source list (from the full energy band) for an observation to the reference source list created using \texttt{IRAF\textbackslash DAOPHOT} from the \CFHTt~image. We set the search radius to $2\arcsec$ and use the WCS from the input (X-ray) source list since the program requires WCS parameters to specify a tangent point for the transform calculations. To eliminate matches with large positional errors from the final transformation calculation, we disabled the $\it{residlim}$ parameter (set to 0), set the $\it{residtype}$ parameter to 0, and set the $\it{residfac}$ parameter to 25. If any residual-to-source pair position error ratio exceeds the value of $\it{residfac}$ it is omitted from the transformation calculation, where the $\it{residtype}$ parameter setting ensures this is completed for each individual source-pair as opposed to averaging all source-pairs. The program then created a transformation matrix with $x$ and $y$ offsets (R.A. and decl.) as well as a rotation and scale parameter. This matrix was then used with $\texttt{wcs\_update}$ to correct the astrometry of the aspect solution file for each individual observation. Since we reprocessed each observation from the level-1 events file, which specifies events in chip coordinates, only the aspect solution file needed to be corrected since it provides the appropriate WCS for $\texttt{acis\_process\_events}$ to convert from chip coordinates to sky coordinates. A complication arose during matching that stems from the exposure mode for many of our observations. The majority of observations were completed in interleaved mode (known as alternating exposure mode), which is carried out by alternating short and long frame times. Two separate event files are produced, one representing events with long frame times and another the short frame times. This is advantageous when observing sources that may be piled-up, such as those in the nucleus of M31. The event files with short frame times all have small livetime exposures and therefore it was difficult to find matching sources for alignment. Since the event files with long frame times always had much larger livetime exposures, they provided more precise astrometric alignment because more sources were detected. Therefore we used the same transformation matrix from the long frame time event file to correct the astrometry for the short frame time event file.

However, 10 ACIS-I and 13 ACIS-S observations were either outside of the PHAT-field or did not have at least three source-pair matches to \CFHTt~image sources. For these observations we first attempted to introduce more matches by using a source list of galaxies from the Sloan Digital Sky Survey (SDSS) Data Release 10. We obtained the source list through the CasJobs batch query service\footnote{\url{http://skyserver.sdss3.org/casjobs/}} because it has no limit on the number of rows output. In the cases where the SDSS match failed, we introduced stars found in the SDSS, whose positions are not necessarily as precise because they may be subject to large proper motions (foreground stars). The proper motions of matched stars that were available were found to be negligible ($<5$ mas yr$^{-1}$). Lastly, we introduced a source list of galaxies from the Two Micron All Sky Survey (2MASS) All-Sky Point Source Catalogue. Since 2MASS publishes an error ellipse, we used $\sqrt{a \times b}$ to determine the error for R.A. and Decl., where $a$ and $b$ are the semi-major and semi-minor axes. The remaining ACIS-I observations were all aligned with the inclusion of the 2MASS point sources. However, 5 ACIS-S observations (314, 4536, 2046, 2047, and 2048) still had $<3$ reliable matches. We then included stars from the Naval Observatory Merged Astrometric Dataset \citep{zacharias11-05}, which also mostly appeared in the SDSS catalogue. This helped us align observation 2048, where we were careful to check that the proper motions of the matched stars were negligible. For the last 4 observations we merged the $\texttt{wavdetect}$ X-ray source lists from the soft, hard, and full energy bands and repeated the matching procedure above with no success.
Since many of our observations overlap one another, we used the astrometric alignments calculated to update the $\texttt{wavdetect}$ X-ray source lists from observations that overlapped the 4 that were still unaligned. This method was still unsuccessful for observation 4536, and so it maintains its original astrometry. The details of astrometric alignment are summarized in Tables \ref{tab:acis-s} and \ref{tab:acis-i}.

\subsection{Merging}

We needed to create merged images of each M31 region (northeast, bulge, southwest) in order to detect the faintest sources. After correcting the astrometry in each individual observation by updating the aspect solution file, we repeated the same preliminary reduction process for events files outlined in Section \ref{sec:datared1} with one minor change: we enabled the energy-dependent subpixel event repositioning algorithm in $\texttt{acis\_process\_events}$ to improve the spatial resolution of point sources on-axis. This algorithm is useful to later distinguish nearby X-ray point sources since most of the observations are centred on the nucleus of M31, and point sources are denser in this region of the galaxy. We then reprojected the cleaned event files for all 29 ACIS-S and 104 ACIS-I observations using $\texttt{reproject\_obs}$. Reprojections were made to the tangent points of the longest observations within the given region.
We then created exposure maps and exposure-corrected images (photons cm$^{-2}$ s$^{-1}$ pixel$^{-1}$) for each observation with the $\texttt{flux\_obs}$ tool, which combined them to create an exposure-corrected image of the 104 ACIS-I observations in the bulge and 5 exposure-corrected images of the various 29 ACIS-S observations based on their overlapping fields (northeast and southwest regions, and 3 regions in the bulge: nucleus and left (northeast) and right (southwest) of the nucleus; see Figure \ref{fig:acis-fov}). The right (southwest) bulge field has 2 ACIS-S regions that we combined due to their proximity. We used a binsize of 1 to maintain the native resolution and weighted spectrum files (corresponding to the soft, hard, and full energy bands) to calculate instrument maps.

We created a total of 18 images throughout the soft, hard, and full energy bands, 3 ACIS-I and 15 ACIS-S, with each representing the regions in Figure \ref{fig:acis-fov}.
The plate scale of the \Chandra\ images is 0.5 arcsecond pixel$^{-1}$, corresponding to 1.9 pc pixel$^{-1}$ at the distance of 776 kpc for M31 adopted at the end of Section \ref{sec:intro}. \Chandra\ ACIS has spatial resolution that ranges from 1\arcsec\ on-axis to 4\arcsec\ at 4\arcmin\ off-axis for 1.5 keV X-rays at 90 per cent encircled energy fraction\footnote{\url{http://cxc.harvard.edu/proposer/POG/}}.
The 3 ACIS-I images of the bulge were each $\approx150$ MB and had dimensions of 5824 pixels by 6145 pixels. Executing $\texttt{flux\_obs}$ and $\texttt{wavdetect}$ to create this image and detect sources required a large amount of memory (e.g.\ $\sim25$ GB for ACIS-I image creation) and storage space ($>1$ TB for all data and ancillary files). We used the Canadian Advanced Network for Astronomical Research (CANFAR; \citealt{gaudet07-11}) to complete this processing.

\subsection{Source Catalogue Creation} \label{sec:catcre}

Source detection was accomplished using $\texttt{wavdetect}$ in order to create a preliminary list of source positions. Starting from our merged exposure-corrected images (one ACIS-I region and five ACIS-S regions) in the soft, hard, and full bands, we used the $\sqrt{2}$ series from 1 to 8 for the $\textit{scales}$ parameter, corresponding exposure maps to reduce false positives, and the inverse of the number of pixels in an image for the $\textit{sigthresh}$ parameter as recommended by the \Chandra\ X-ray Center. Since our merged images are larger than the standard $1024\times1024$ pixels for most images, the $\textit{sigthresh}$ parameter needed to be modified to reduce the number of false sources detected. It was set to $3\times10^{-8}$ for the ACIS-I image and $\sim$few$\times10^{-6}$ for the five ACIS-S images. The $\textit{expthresh}$ parameter was changed from its default value of 0.1 to 0.001 for all 6 merged images. Pixels with a relative exposure (pixel exposure value over maximum value of exposure in the map) less than $\textit{expthresh}$ are not analysed, and adopting the default value would exclude the analysis of many of the low-exposure regions in our merged images. All other $\texttt{wavdetect}$ parameters were left at their default values. The source lists we obtained were then combined into a master source list using the $\texttt{match\_xy}$ tool from the Tools for ACIS Review and Analysis (TARA) package\footnote{\url{http://www2.astro.psu.edu/xray/docs/TARA/}}, resulting in a candidate list of 1068 sources. In addition, to ensure all possible sources were detected $\texttt{reproject\_image}$ was used to merge the 3 ACIS-S bulge region counts images with the ACIS-I counts image. Running $\texttt{wavdetect}$ on each energy range for this master bulge region image and combining the source lists following the same procedure as above we recovered 331 additional sources (unique from the original 1068). Therefore we had a total of 1399 $\texttt{wavdetect}$ sources.

To obtain source properties from our preliminary list of source positions we used {\em ACIS Extract} \citep[{\em AE};][]{broos05-10}, which performed source extraction and characterization. AE analyses the level-2 event files of each observation individually before merging and determining source properties. In order to create a reliable catalogue, we followed the methods outlined in the validation procedure\footnote{\label{foot:valid}\url{http://www2.astro.psu.edu/xray/docs/TARA/ae\_users\_guide/procedures/}} used by the authors of AE. This multi-step process (summarized in Figure 1 of \citealt{broos05-10}) involved many iterations of extracting, pruning, and repositioning sources in the candidate catalogue. We pruned sources using the {\em AE} parameter $pns$ (`$prob\_no\_source$' or the p-value for no-source hypothesis), which calculates the Poisson probability of a detection not being a source by taking the uncertainty of the local background into account. Using the $pns$ value removes the bias inherent in using the traditional $3\sigma$ source significance criterion, where sources with low count rates would have been left out.

Before pruning any source we visually inspected it in any of the observations it appeared to be sure that it was insignificant. In some cases 2 neighbouring sources were both selected for pruning, but it was often found that either one would survive if the other was removed. The most advantageous aspect of the validation procedure was the ability to review each source and select the most likely position based on the properties of an individual source. Specifically, the repositioning stage displays the original catalogue position and also calculates 3 other position estimates: the mean data, correlation, and maximum-likelihood reconstruction positions. The catalogue and/or centroid positions were used most often, but for sources with large off-axis angles ($\gtrsim5\arcmin$) the correlation position was used, whereas for crowded sources with overlapping PSFs the maximum-likelihood reconstruction position was used (see \S{7.1} of \citealt{broos05-10} for a detailed description of source positions). {\em AE} also produces smoothed residual images for each observation, which are the smoothed residuals remaining after the point source models are subtracted from the observation data. The residual image is scaled to emphasize only bright residuals, which are possible sources that may have been missed by $\texttt{wavdetect}$ or were accidentally pruned. Inspecting the observed events and point source model for a bright residual reveals whether it is an artefact or likely point source. We found \rsrc\ sources using the residual images that were added to our source list. After catalogue positions were validated, the one-pass photometry procedure (follows the validation procedure, see footnote \ref{foot:valid}) completed the final extraction for all \num\ sources in our final M31 catalogue; output source properties are summarized in Tables \ref{tab:srclist}$-$\ref{tab:srclist2}. To be included in the final catalogue a source was required to have a $pns$ value $\leq1\times10^{-2}$ (default in {\em AE}) in any of the energy bands (full, soft, or hard). We chose this value since it was used for the \Chandra\ Carina Complex Project \citep{broos05-11} to balance sensitivity with source detection significance. In the full band, all \fbandsrces\ (\fullbandpct\%) of our sources had a $pns$ value $\leq1\times10^{-2}$.

\section{X-ray Source Catalogue Properties}

\subsection{Source Catalogue} \label{sec:src-info}

\begin{figure*}
\begin{tabular}{cc}
\includegraphics[width=1.0\columnwidth]{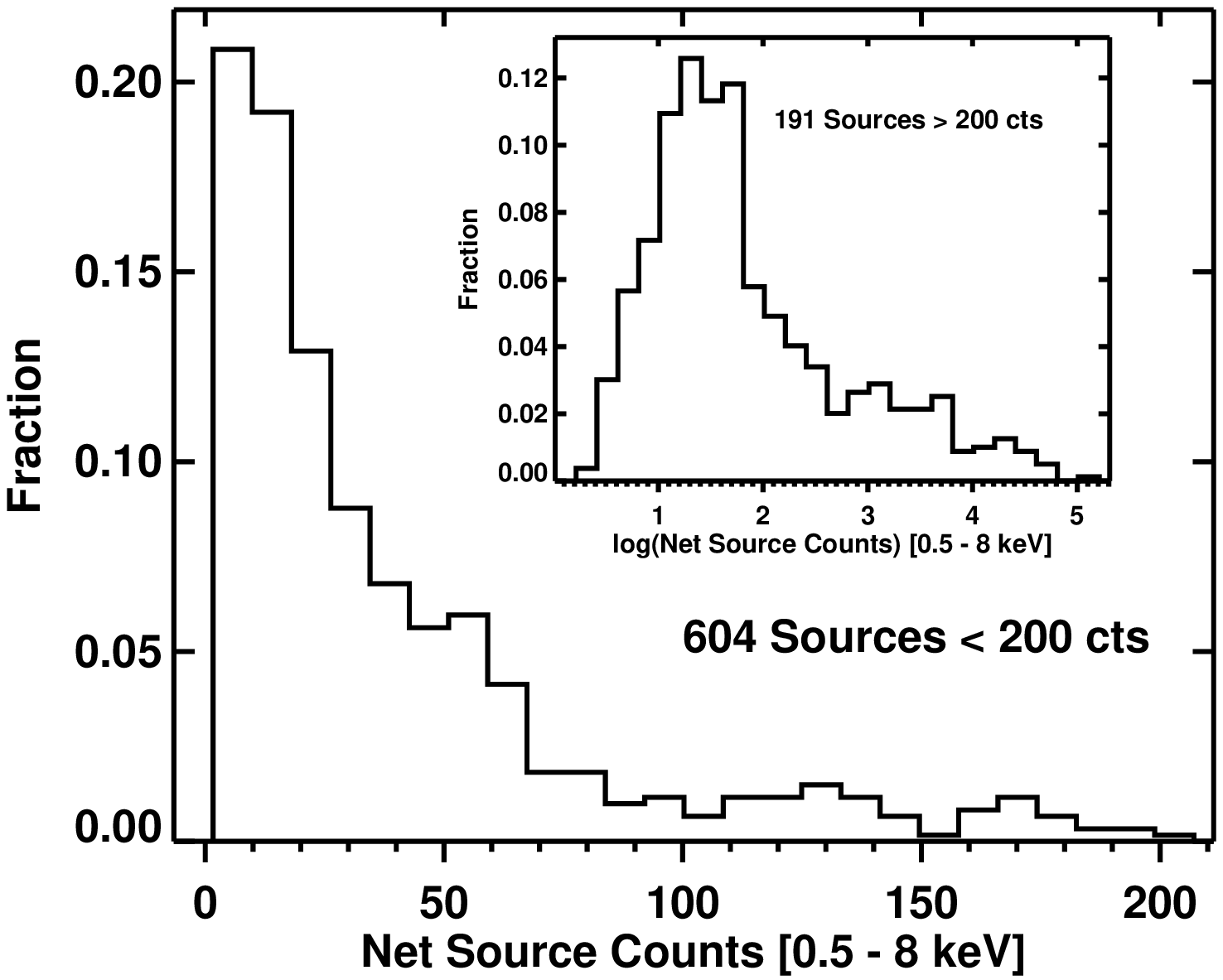}
\includegraphics[width=1.0\columnwidth]{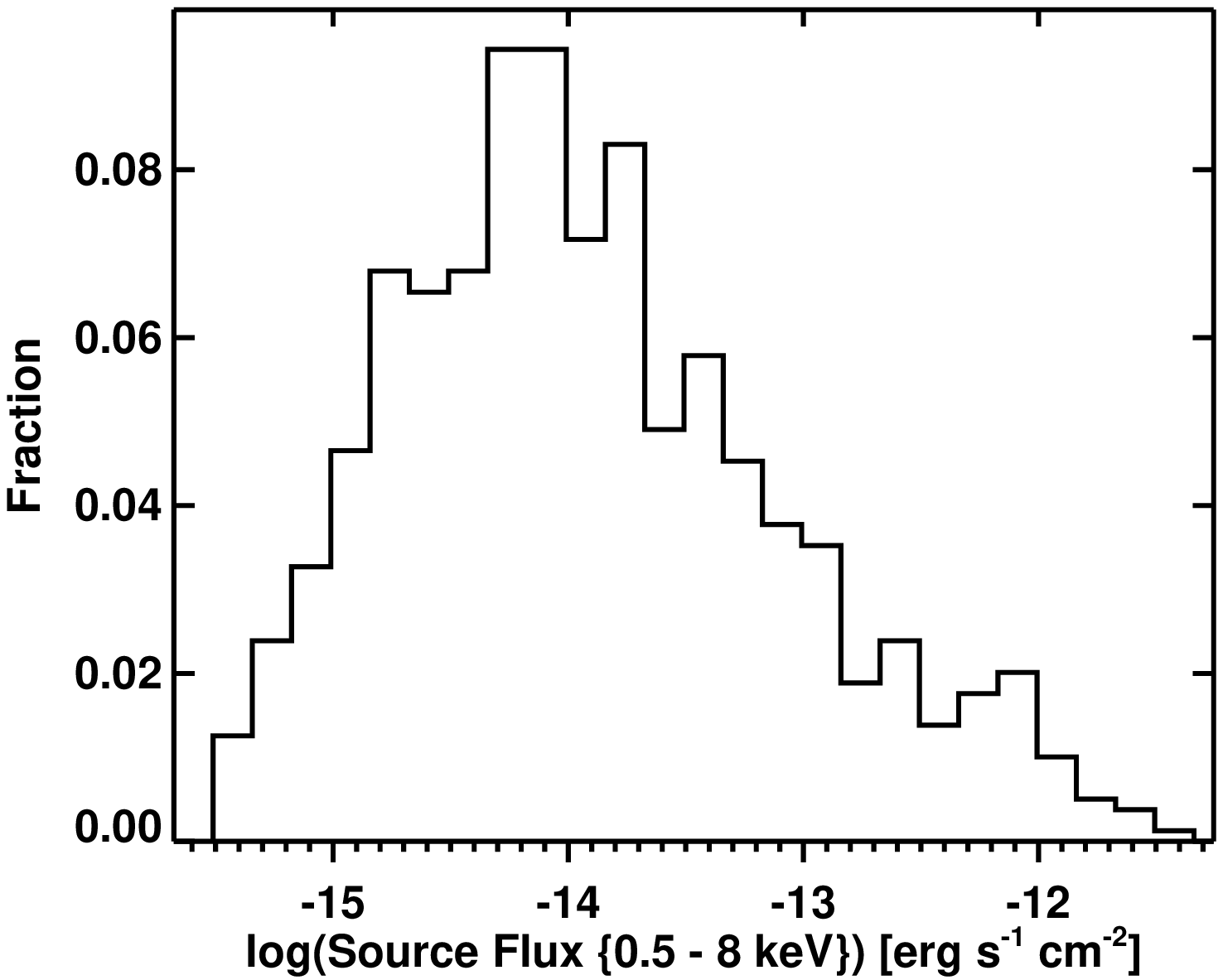}
\end{tabular}
\caption{Histograms of the net source counts (left panel) and source flux (right panel) in the full energy band. The left panel shows sources with $<200$ counts while the inset shows a histogram of all \num\ sources. While the majority of sources have $<200$ counts, the 191 sources (mostly in the nucleus) with $\gtrsim$200 counts are candidates for spectral modeling. In the right panel, the peak in the distribution is near our 90\% detection limit of $\approx5\times10^{-15}$ \es\ cm$^{-2}$ ($3\times10^{35}$ \es).}\label{fig:hists}
\end{figure*}

Our final M31 catalogue consists of \num\ X-ray point sources. Their properties are summarized in Tables \ref{tab:srclist}$-$\ref{tab:srclist2}. The detailed FITS tables from {\em AE}, which include $\sim100$ other source photometric properties used in {\em AE} (e.g.\ background region metrics) in 16 energy bands (see Section 7.8 of the {\em AE} manual for these bands), are available on the journal website.
Of the \num\ sources, \nes\ are in the northeast portion of M31, \blg\ are in the bulge, and \sw\ are in the southwest. The fluxes in Table \ref{tab:srclist2} were calculated using conversion factors of 5.47, 1.86, and 3.37 in units of $10^{-9}$ erg photons$^{-1}$ for the full, soft, and hard bands (to convert from photon flux in photons cm$^{-2}$ s$^{-1}$ to energy flux in erg cm$^{-2}$ s$^{-1}$). Photon fluxes were estimated by {\em AE} ($\emph{flux2}$ parameter) based on the number of net source counts ($net\_cts$), the exposure time, and the mean ancillary response function ($MEAN\_ARF$) in the given energy band. This flux estimate suffers from a systematic error (compared to the true incident flux) because the $MEAN\_ARF$ is the correct effective area normalization based on an incident spectrum that is flat. To obtain a more accurate flux estimate, one should sum the flux values over narrow energy bands. We performed this summation for the full, soft, and hard energy bands using the fluxes in narrow energy ranges provided by {\em AE}: $0.5-1.0$, $1.0-2.0$, $2.0-4.0$, $4.0-6.0$, and $6.0-8.0$. We found a difference of $-8$\%, 5\%, and 13\% between the summed flux values and those from the $\emph{flux2}$ parameter. This percent difference is much smaller than the expected uncertainties for fluxes based on the Poisson errors for the net counts. We reported the summed fluxes in the full, soft, and hard bands in the last 3 columns of Table \ref{tab:srclist2}. We used the {\em AE} $\emph{flux2}$ fluxes for our analysis. We converted the photon flux into an energy flux assuming an absorbed power-law spectrum with $\Gamma=1.7$ and $N_{H} = 6.66\times10^{20}$ cm$^{-2}$ (details in Section \ref{sec:sens}). In Figure \ref{fig:hists}, we show histograms of the net counts (left panel) and source flux in erg s$^{-1}$ cm$^{-2}$ (right panel) for all our sources. The left panel shows sources with $<200$ counts, while the inset shows a log distribution of all sources, clearly indicating the majority of sources have $<100$ counts. The 191 sources that have $\gtrsim200$ counts are suitable for spectral modelling. The source flux histogram shows a peak in the flux distribution near our 90\% detection limit of $\approx5\times10^{-15}$ \es\ cm$^{-2}$ ($3\times10^{35}$ \es). The X-ray point source population of M31 is comprised of LMXBs and HMXBs, supernova remnants, and background AGNs. To determine the makeup of our population of point sources we created XLFs and used various catalogues from previous studies.

\begin{table*}
\scriptsize
\caption{M31 Source List \label{tab:srclist}}
\resizebox{\textwidth}{!}{%

\begin{tabular}{cccccccccccccccc}
\hline\hline
Source	&	CXOU J	&	R.A. (J2000)	&	Decl. (J2000)	&	Distance	&	PosErr	&	$\theta$	&	No. of	&	Detector	&	Region	&	Tot Exp	&	Tot Exp. Map	&	R$_{src}$	&	SNR	&	$E_{\rm{median}}$	&	Match	\\
No. & & ($\degr$)	&($\degr$)	&	($\arcmin$) 	&	($\arcsec$)	&	($\arcmin$) & Obs	&	&	&	(ks) & Value (s cm$^{2}$)	&	(sky pixel)	&	&	(keV)	&	\vspace{ .1cm}	\\	
(1) &	(2) & (3)	&	(4)	&	(5)	&	(6) & (7) & (8)	&	(9)	&	(10)	&	(11)	&	(12)	&	(13)	&(14)	&	(15)	&	(16)	\\	
\hline
        1 & 004542.90+414312.6 & 11.428779 & 41.720189 & 43.03 & 0.2 & 4.1 & 2 & ACIS-S & NE & 49 & 1.14E+07 & 4.5 & 6.5 & 2.1 & 
 \\ 2 & 004551.05+414452.4 & 11.462750 & 41.747912 & 45.26 & 0.3 & 3.5 & 2 & ACIS-S & NE & 49 & 1.64E+07 & 3.5 & 2.7 & 2.7 & 
 \\ 3 & 004551.30+414220.7 & 11.463769 & 41.705754 & 43.74 & 0.2 & 2.8 & 3 & ACIS-S & NE & 64 & 2.20E+07 & 2.8 & 2.4 & 2.3 & AGN
 \\ 4 & 004552.93+414441.8 & 11.470551 & 41.744965 & 45.42 & 0.2 & 3.1 & 2 & ACIS-S & NE & 49 & 1.68E+07 & 3.0 & 2.8 & 2.0 & 
 \\ 5 & 004555.72+414551.8 & 11.482172 & 41.764389 & 46.56 & 0.3 & 3.7 & 2 & ACIS-S & NE & 49 & 1.67E+07 & 4.0 & 2.5 & 2.5 & 
 \\ 6 & 004556.82+414440.8 & 11.486787 & 41.744673 & 45.98 & 0.2 & 2.6 & 2 & ACIS-S & NE & 49 & 1.72E+07 & 2.5 & 2.5 & 2.7 & 
 \\ 7 & 004556.99+414831.7 & 11.487497 & 41.808829 & 48.48 & 0.2 & 6.2 & 2 & ACIS-S & NE & 49 & 1.60E+07 & 9.3 & 9.6 & 2.4 & AGN
 \\ 8 & 004559.07+414113.0 & 11.496132 & 41.686945 & 44.27 & 0.2 & 2.6 & 5 & ACIS-S & NE & 91 & 3.38E+07 & 3.0 & 5.3 & 2.9 & 
 \\ 9 & 004602.43+414515.7 & 11.510137 & 41.754377 & 47.16 & 0.2 & 3.1 & 3 & ACIS-S & NE & 63 & 1.84E+07 & 3.3 & 5.7 & 2.1 & 
 \\ 10 & 004602.70+413856.7 & 11.511251 & 41.649095 & 43.61 & 0.3 & 3.4 & 3 & ACIS-S & NE & 41 & 1.62E+07 & 4.1 & 3.6 & 1.4 & 
 \\ 
\hline
\end{tabular}
}
\begin{list}{}{}
\setlength\labelwidth{0cm}
\item $\bf{Notes}$. Column 2: source ID, which contains the source coordinates (J2000.0). Column 5: distance in arcminutes of the source from the centre of M31 (J004244.33+411607.50). Column 6: positional uncertainty $\sqrt{\sigma_{x}^{2}+\sigma_{y}^{2}}$, where the single-axis position errors $\sigma_{x}$ and $\sigma_{y}$ are estimated from the standard deviations of the PSF in the extraction region and the number of counts extracted. Column 7: average off-axis angle for merged observations. Column 8: number of observations extracted. Column 9: source detected in ACIS-I, ACIS-S, or Both. Column 10: for a source detected in ACIS-S or Both, indicates which region from Figure \ref{fig:acis-fov} it belongs to. Columns 11 \& 12: total values for merged observations. Column 13: average radius of the source extraction region (1 sky pixel = $0.492\arcsec$). Column 14: photometric significance (net counts / upper error on net counts) ($0.3-8.0$ keV). Column 15: background-corrected median photon energy ($0.3-8.0$ keV). Column 16: cross-match results from Section \ref{sec:lognlogs}: active galactic nuclei (AGN) or low-mass X-ray binary (LMXB). \\
(This table is published in its entirety in the electronic edition of the journal. A portion is shown here for guidance regarding its form and content.)
\end{list}
\end{table*}

\begin{table*}
\scriptsize
\caption{Additional M31 Source Properties \label{tab:srclist2}}
\resizebox{\textwidth}{!}{%

\begin{tabular}{cccccccccccccccccccccc}
\hline\hline
Source	&	$pns$	&	$pns$	&	$pns$	&	$net\_cts$ 	&	$\sigma_{\rm{upper}}$	&	$\sigma_{\rm{lower}}$	&	$net\_cts$	&	$\sigma_{\rm{upper}}$	&	$\sigma_{\rm{lower}}$	&	$net\_cts$	&	$\sigma_{\rm{upper}}$	&	$\sigma_{\rm{lower}}$	&	$flux$	&	$flux $	&	$flux $	&	Summed $flux$	&	Summed $flux $	&	Summed $flux $	\\
No. & ($0.5-8.0$ keV)	&($0.5-2.0$ keV)	&	($2.0-8.0$ keV)	&	\multicolumn{3}{c}{($0.5-8.0$ keV)} 	& \multicolumn{3}{c}{($0.5-2.0$ keV)}	&	\multicolumn{3}{c}{($2.0-8.0$ keV)} & ($0.5-8.0$ keV)	&($0.5-2.0$ keV)	&	($2.0-8.0$ keV)		& ($0.5-8.0$ keV)	&($0.5-2.0$ keV)	&	($2.0-8.0$ keV)		\\	
\cmidrule(lr){5-7} \cmidrule(lr){8-10} \cmidrule(lr){11-13}
(1) &	(2) & (3)	&	(4)	&	(5)	&	(6) & (7) & (8)	&	(9)	&	(10)	&	(11)	&	(12)	&	(13)	&	(14)	&	(15)	&	(16)	&	(17)	&	(18)	&	(19)	\\	
\hline
        1 & 0.00E+00 & 1.40E-45 & 1.37E-23 & 55.65 & 8.60 & 7.53 & 33.52 & 6.90 & 5.80 & 22.13 & 5.87 & 4.76 & 3.07E-14 & 3.97E-15 & 8.83E-15 & 2.24E-14 & 4.06E-15 & 6.43E-15
 \\ 2 & 7.48E-12 & 6.96E-09 & 6.96E-05 & 13.01 & 4.84 & 3.69 & 7.65 & 3.96 & 2.76 & 5.35 & 3.60 & 2.38 & 5.37E-15 & 6.33E-16 & 1.65E-15 & 5.64E-15 & 6.45E-16 & 2.31E-15
 \\ 3 & 2.17E-09 & 1.18E-08 & 5.55E-03 & 10.96 & 4.57 & 3.41 & 7.63 & 3.96 & 2.76 & 3.33 & 3.18 & 1.91 & 3.40E-15 & 4.76E-16 & 7.70E-16 & 3.02E-15 & 4.83E-16 & 9.89E-16
 \\ 4 & 4.98E-14 & 1.50E-10 & 1.19E-05 & 13.32 & 4.84 & 3.69 & 7.79 & 3.96 & 2.76 & 5.53 & 3.60 & 2.38 & 5.46E-15 & 6.34E-16 & 1.70E-15 & 3.84E-15 & 5.92E-16 & 1.30E-15
 \\ 5 & 2.19E-09 & 1.08E-06 & 2.45E-04 & 11.71 & 4.71 & 3.56 & 6.53 & 3.78 & 2.58 & 5.18 & 3.60 & 2.38 & 4.86E-15 & 5.34E-16 & 1.62E-15 & 4.46E-15 & 6.04E-16 & 1.65E-15
 \\ 6 & 1.04E-13 & 2.88E-08 & 5.85E-07 & 11.56 & 4.57 & 3.41 & 5.84 & 3.60 & 2.37 & 5.72 & 3.60 & 2.37 & 4.66E-15 & 4.66E-16 & 1.74E-15 & 4.33E-15 & 4.98E-16 & 1.77E-15
 \\ 7 & 0.00E+00 & 2.01E-41 & 0.00E+00 & 117.40 & 12.23 & 11.18 & 50.47 & 8.34 & 7.26 & 66.93 & 9.55 & 8.48 & 5.10E-14 & 4.33E-15 & 2.19E-14 & 3.73E-14 & 4.19E-15 & 1.54E-14
 \\ 8 & 1.88E-39 & 1.19E-12 & 2.21E-28 & 40.21 & 7.54 & 6.45 & 12.37 & 4.71 & 3.56 & 27.84 & 6.46 & 5.35 & 8.61E-15 & 5.19E-16 & 4.51E-15 & 6.79E-15 & 4.64E-16 & 3.34E-15
 \\ 9 & 0.00E+00 & 6.44E-38 & 1.24E-16 & 44.71 & 7.84 & 6.76 & 28.51 & 6.45 & 5.35 & 16.20 & 5.21 & 4.08 & 1.69E-14 & 2.12E-15 & 4.63E-15 & 1.21E-14 & 2.04E-15 & 3.74E-15
 \\ 10 & 3.91E-18 & 8.90E-20 & 5.38E-03 & 21.35 & 5.88 & 4.76 & 17.42 & 5.33 & 4.20 & 3.93 & 3.40 & 2.16 & 1.01E-14 & 1.58E-15 & 1.43E-15 & 7.11E-15 & 1.76E-15 & 1.20E-15
 \\
\hline
\end{tabular}
}
\begin{list}{}{}
\setlength\labelwidth{0cm}
\item $\bf{Notes}$. Columns (2)-(4) represent the $pns$ values, which are the Poisson probability of a detection not being a source (discussed in Section \ref{sec:catcre}). Columns (5)-(13) are the net counts in a given energy range with 90\% upper and lower uncertainty limits. Columns (14)-(16) show the fluxes (the $\emph{flux2}$ parameter in {\em AE}) in units of erg cm$^{-2}$ s$^{-1}$. Columns (17)-(19) are the fluxes in erg cm$^{-2}$ s$^{-1}$ calculated by summing narrow energy bands and thus avoiding systematic errors in the $\emph{flux2}$ parameter from {\em AE} (see Section \ref{sec:src-info}). Conversion factors were 5.47, 1.86, and 3.37 in units of $10^{-9}$ erg photons$^{-1}$ for the full, soft, and hard bands (to convert from photon flux in photons cm$^{-2}$ s$^{-1}$ to energy flux in erg cm$^{-2}$ s$^{-1}$). The conversion factors account for foreground absorption using an absorbed power-law spectrum with $\Gamma=1.7$ and $N_{H} = 6.66\times10^{20}$ cm$^{-2}$. To convert energy fluxes to luminosity in erg s$^{-1}$ multiply by $7.2\times10^{49} \left(\frac{d}{776 \rm{\, kpc}}\right)^{2}$ cm$^{2}$. For some sources the soft or hard band had $<0$ counts, and so uncertainties could not be determined and are represented as -99.99. By extension, some fluxes in the soft or hard band were $<0$ and so luminosities appear as -9.99. Each source has a $pns$ value $<1\times10^{-2}$ in at least one energy band. \\
(This table is published in its entirety in the electronic edition of the journal. A portion is shown here for guidance regarding its form and content.)
\end{list}
\end{table*}

\subsection{Cross-Correlation With Existing Catalogues} \label{sec:xmm}

\begin{figure}
\includegraphics[width=1.0\columnwidth]{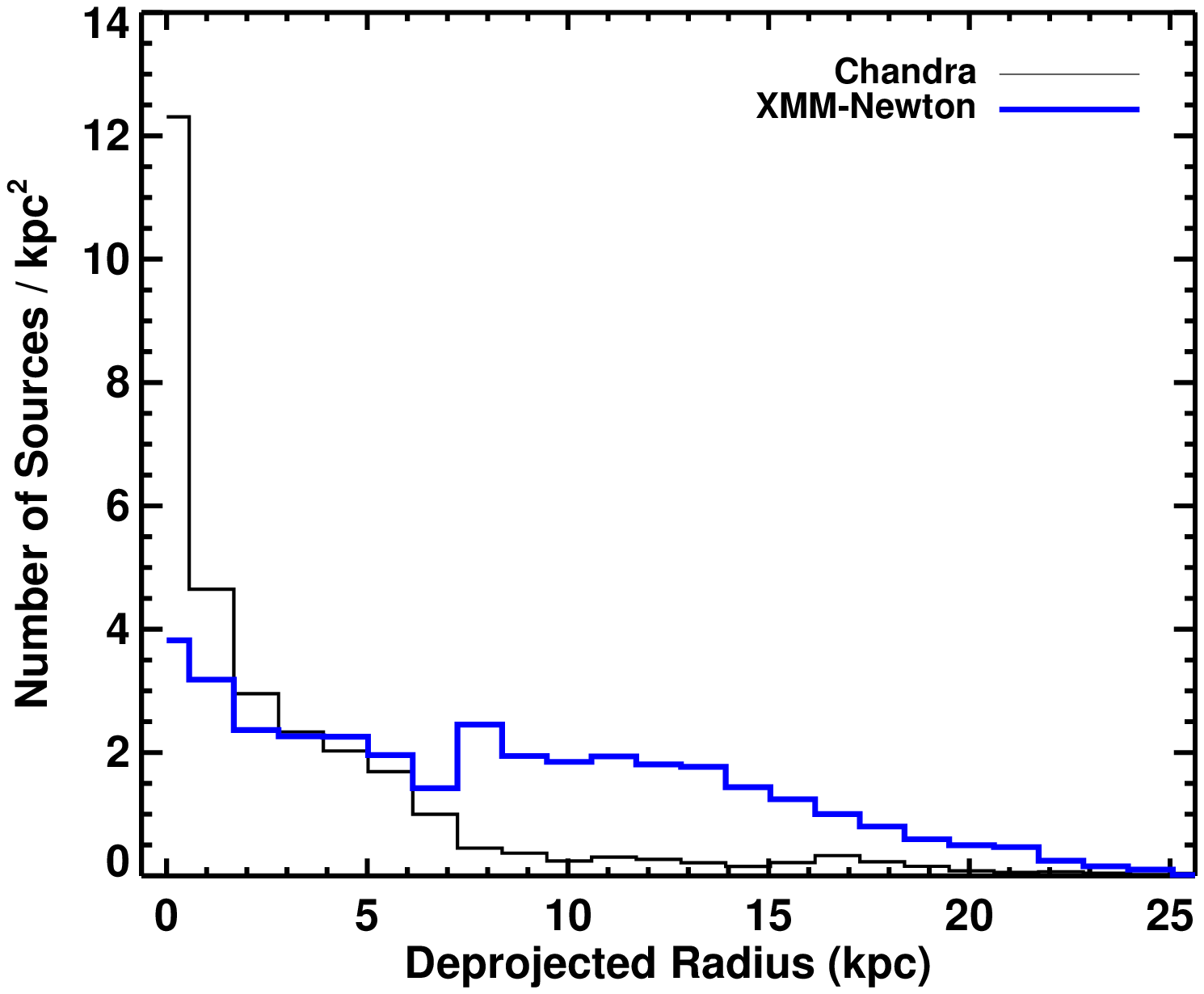}
\caption{Deprojected radial source distribution of X-ray point sources in our catalogue (black) compared with those from the \xmmn\ survey (blue). The density of our \Chandra\ sources drops off considerably around a few kpc because the coverage of our survey is sparse (see Figure \ref{fig:acis-fov}). However, the deeper observations of the nucleus combined with the sub-arcsecond resolution of \Chandra\ results in a $\sim3$-fold increase in the number of sources within the central kpc.}\label{fig:raddist}
\end{figure}

Of the many previous studies in M31 summarized in Tables \ref{tab:sum} - \ref{tab:chandra}, not all actually publish a traditional X-ray catalogue. Approximately half the studies were focused on only one specific X-ray population. Our first comparisons are to the \xmmn\ catalogue \citep{stiele10-11} because it is complete throughout M31 to a limiting luminosity of $\sim10^{35}$ \es. Of the 1948 X-ray sources identified, only 979 appear within the field of view of our \Chandra\ data. Figure \ref{fig:raddist} shows the deprojected radial distribution of X-ray sources in our catalogue (black) compared to those from the \xmmn\ survey (blue). Because our survey has sporadic coverage, only the nuclear region, where the exposure is significant, do we see a greater source density. Also, the subarcsecond resolution of \Chandra\ allowed us to separate closely-spaced sources that \xmmn\ was unable to resolve due to its 5\arcsec\ PSF.

\begin{figure*}
\begin{tabular}{cc}
\includegraphics[width=1.0\columnwidth]{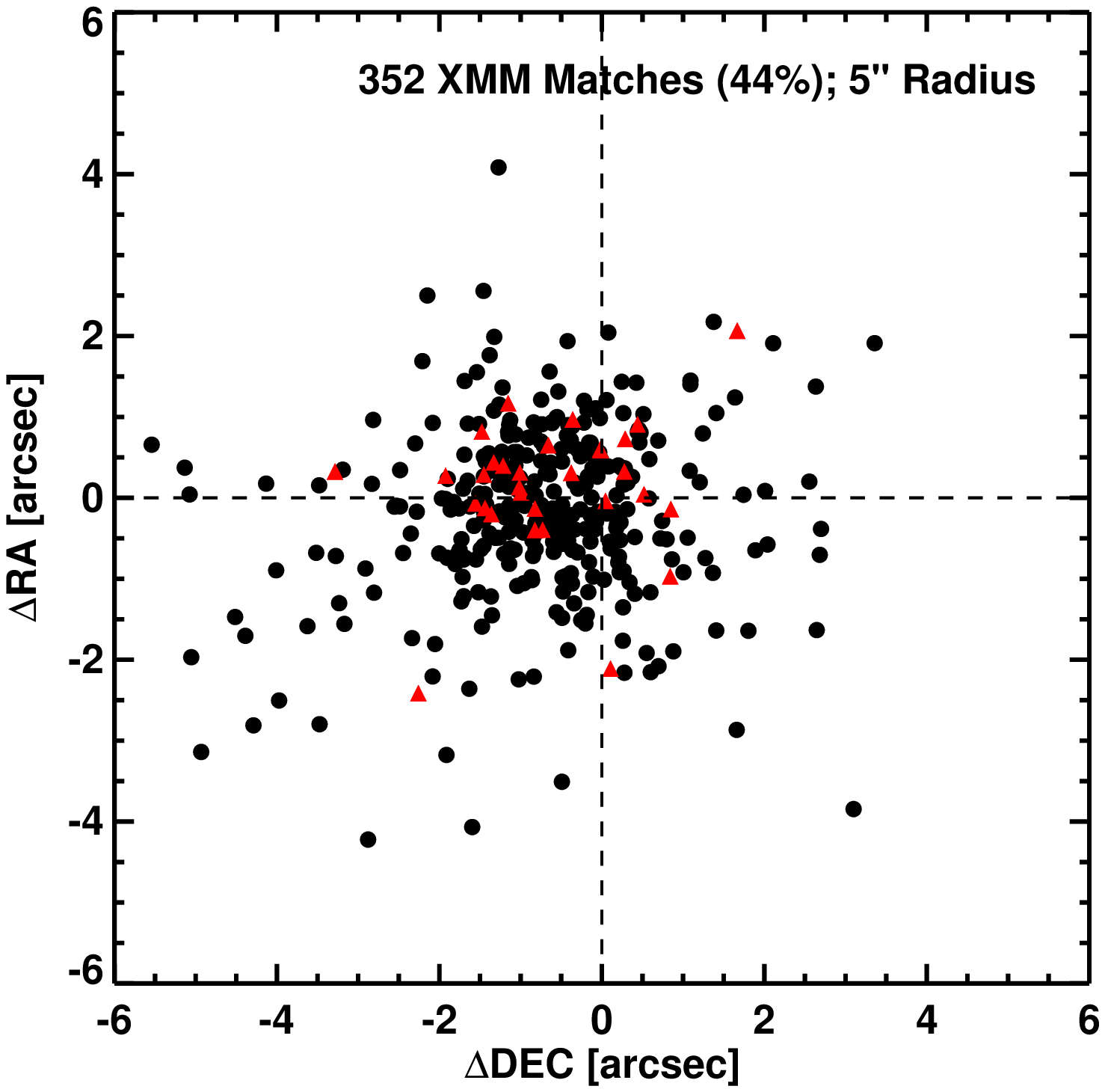}
\includegraphics[width=1.0\columnwidth]{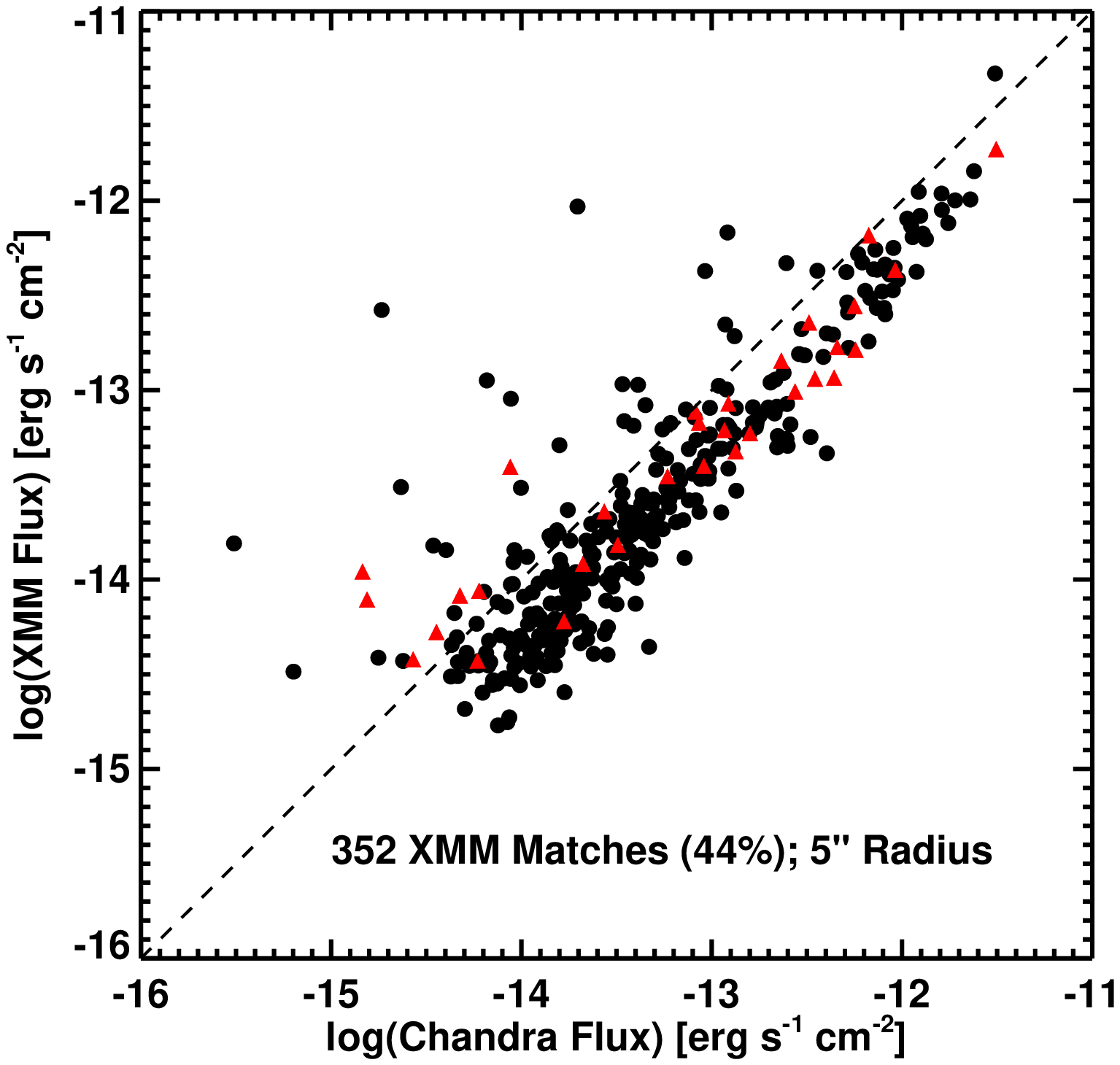}
\end{tabular}
\caption{A comparison of the \xcmatchuniq\ (\matchper\%) unique \Chandra\ X-ray sources (a total of \xcmatch\ (\matchperall\%) were matched) detected in this work that were the closest matches to those from the \xmmn\ catalogue within 5\arcsec\ (approximately the positional uncertainty associated with \xmmn\ sources). The left panel shows the approximate positional offset between the matches while the right panel compares the fluxes. The red triangles represent \xmmn\ sources that were matched to multiple \Chandra\ sources, where the values plotted came from the closest-matched source. \xmmn\ fluxes ($0.2-4.5$ keV) have been corrected to the \Chandra\ energy range ($0.5-8.0$ keV) assuming the spectral model from Section \ref{sec:sens}. \Chandra\ fluxes are $\sim50\%$ brighter, which is due to either cross-calibration between the observatories or the assumed conversion factor used for \xmmn\ fluxes.}\label{fig:xmmmatch}
\end{figure*}

We also matched our catalogue to the \xmmn\ catalogue using a matching radius of 5\arcsec, equivalent to \xmmn's source positional uncertainty. We found that \xcmatchuniq\ \xmmn\ sources matched to \xcmatch\ (\matchperall\%) of our \Chandra\ sources, where the \xcmatchuniq\ unique \Chandra\ sources (closest matches) made up \matchper\ per cent of our catalogue. Because the \xmmn\ PSF is much larger than \Chandras, multiple \Chandra\ sources were matched to within 5\arcsec\ of an \xmmn\ source. The median offset between matches was \medsepx\arcsec. In Figure \ref{fig:xmmmatch} we show the positional offset between the matches (left panel) and compare the average fluxes from both surveys (right panel). \Chandra\ fluxes appeared systematically brighter with a mean value of $\sim50\%$, either due to cross-calibration between the observatories or the assumed conversion factor used for \xmmn\ fluxes. To address the sources in the \xmmn\ catalogue that were not matched to our catalogue, we compiled a list of all the original input sources throughout all three fields (1399) and included an additional $\sim5000$ sources from a low-significance $\texttt{wavdetect}$ run. Matching this list to the \xmmn\ catalogue within 5\arcsec\ we find that \xmmxtramat\ (\xmmxtramatpct\%) of \xmmn\ sources were matched. Therefore many sources can be identified as candidates in the \Chandra\ observations but not `detected' due to a low significance (e.g., shallow exposure, transience).

Our catalogue was also matched to within 1\arcsec\ of previous \Chandra\ catalogues of various X-ray sources. From the \Chandra\ surveys in Table \ref{tab:chandra}, a total of \allprevc\ sources (some authors omitted sources from their catalogue if they were not new) were cross-matched with those from our catalogue. Among the past \Chandra\ catalogues there is some redundancy as groups do not always exclude previously detected sources from their catalogue.
Using this matching radius we find only \catc\ of our sources that matched to previously detected \Chandra\ sources, with a median offset of \medsepc\arcsec. \newc\ of our catalogue sources were detected for the first time from \Chandra\ observations. When combining the unique matches from both \xmmn\ and previous \Chandra\ surveys we find that \alluniq\ matched from our catalogue of \num\ sources, meaning \absnew\ of our X-ray sources were detected for the first time. These new sources have a factor of 2 smaller exposure times, a factor of three fewer counts and photon fluxes, a factor of six fewer observations per source, and $pns$ values that are twelve orders of magnitude less significant, when compared to median values of the whole catalogue (e.g.\ $pns\approx10^{-17}$). When compared to only the matched sources, these differences are exacerbated even more so. Matching results are shown in Table \ref{tab:matches}. The cross-match information between our catalogue and previous \xmmn\ and \Chandra\ catalogues, including their identification numbers/values (where available), is shown in Table \ref{tab:crossmatch}. We also show the number of our catalogue sources matched for each source classification type from other catalogues in Table \ref{tab:totcross}.

\begin{table}
\caption{Matching Results\label{tab:matches}}	
\resizebox{\columnwidth}{!}{%

\begin{tabular}{@{}ccccc@{}}	
\hline\hline
	&	\xmmn	&	Previous \Chandra	&	Total Matched	&	New sources	\\
\hline
This work (\num) &	\xcmatch\ (\xcmatchuniq\ unique) & 	\catc	&	\alluniq	&	\absnew	\\
\hline
\end{tabular}
}
\begin{list}{}{}
\item A 5\arcsec\ radius was used for matching to \xmmn, while a 1\arcsec\ radius was used when matching to previous \Chandra\ catalogues. \xcmatch\ of our catalogue sources were matched to \xcmatchuniq\ \xmmn\ sources.
\end{list}
\end{table}

\begin{table*}
\scriptsize
\caption{X-ray Catalogue Cross-Matches \label{tab:crossmatch}}
\resizebox{\textwidth}{!}{%

\begin{tabular}{ccccccccccccc}
\hline\hline
Source	&	CXOU J	&	\multicolumn{2}{c}{\xmmn\ Match}	&	\multicolumn{2}{c}{\Chandra\ Match} 	&	\multicolumn{3}{c}{LMXB Match}	&	\multicolumn{4}{c}{AGN Match}		\\
\cmidrule(lr){3-4} \cmidrule(lr){5-6} \cmidrule(lr){7-9} \cmidrule(lr){10-13}
No. &	 &	ID	& Classification		&	Catalogue	&	ID 	& \citeauthor{peacock10-10}; GC	&	\citeauthor{stiele10-11}; Field & \citeauthor{stiele10-11}; GC	&	PHAT	&	SDSS DR12	&	NED		&	SIMBAD		\\	
(1) &	(2) & (3)	&	(4)	&	(5)	&	(6) & (7) & (8)	&	(9)	&	(10)&	(11)&	(12)&	(13)	\\	
\hline
1 & 004542.90+414312.6 & 1685 & $<$hard$>$ & - & - & - & - & - & - & - & - & -
 \\ 2 & 004551.05+414452.4 & - & - & - & - & - & - & - & - & - & - & -
 \\ 3 & 004551.30+414220.7 & - & - & - & - & - & - & - & 3900 & - & - & -
 \\ 4 & 004552.93+414441.8 & - & - & - & - & - & - & - & - & - & - & -
 \\ 5 & 004555.72+414551.8 & - & - & - & - & - & - & - & - & - & - & -
 \\ 6 & 004556.82+414440.8 & - & - & - & - & - & - & - & - & - & - & -
 \\ 7 & 004556.99+414831.7 & 1716 & $<$hard$>$ & - & - & - & - & - & 1938 & - & - & -
 \\ 8 & 004559.07+414113.0 & - & - & - & - & - & - & - & - & - & - & -
 \\ 9 & 004602.43+414515.7 & 1732 & $<$SNR$>$ & - & - & - & - & - & - & - & - & -
 \\ 10 & 004602.70+413856.7 & - & - & - & - & - & - & - & - & - & - & -
\\
\hline
\end{tabular}
}
\begin{list}{}{}
\setlength\labelwidth{0cm}
\item $\bf{Notes}$. 
This table summarizes the details of the cross-match between our catalogue and various others. Columns (1) and (2) represent our catalogue. Columns (3) and (4) are the \xmmn\ catalogue identification number and classification. Column (5) is the \Chandra\ catalogue matched to: BA \citep{barnard01-14}, DS02 \citep{di-stefano05-02}, DS04 \citep{di-stefano07-042}, HO \citep{hofmann07-13}, KA \citep{kaaret10-02}, KO \citep{kong10-02}, VO \citep{voss06-07}, WI \citep{williams07-04}. Column (6) is the catalogue source identification value taken from each respective paper. Columns (7)-(9) represent matches to LMXBs from \citet{peacock10-10} globular clusters (GC), and \citet{stiele10-11} field and globular clusters (GC), with the corresponding names or identification number. Columns (10)-(13) show the results of AGN matching to various catalogues. For PHAT, we used the Andromeda project identification number from \citet{johnson04-15}. See Section \ref{sec:lognlogs} for more details on matching. \\
(This table is published in its entirety in the electronic edition of the journal. A portion is shown here for guidance regarding its form and content.)
\end{list}
\end{table*}

\begin{table}
\caption{Summary of X-ray Catalogue Cross-Matches\label{tab:totcross}}	
\resizebox{\columnwidth}{!}{%

\begin{tabular}{@{}ccccc@{}}	
\hline\hline
Source Classification	&	Number Matched	&	Number Matched (including candidates)	\\
\hline
AGN		&	29	&	40	\\
XRB		&	8	&	33	\\
Globular Cluster (LMXB)	&	46	&	54	\\ 
Supernova Remnant (SNR)		&	12	&	14	\\
Foreground Star	&	6	&	29	\\
\hline
\end{tabular}
}
\begin{list}{}{}
\item Total number of sources from our catalogue matched to previously identified sources from catalogues listed in Table \ref{tab:crossmatch}. The last column includes the sources from \citet{stiele10-11} that were identified as candidates in addition to the confirmed sources (second column) from all other catalogues.
\end{list}
\end{table}

\subsection{Sensitivity Curve} \label{sec:sens}

In order to create corrected XLFs (log$N$-log$S$) we had to evaluate the sensitivity of each point in the survey field, which gives the energy flux at which a source would be detected. Complications arise when attempting to compute sensitivity for overlapping observations/regions, which are prevalent in our survey field. In addition, the \CIAO\ tools are not designed for this type of analysis and also do not accommodate combining ACIS-I/S observations. Therefore we take a statistical approach and follow the method of \citet{georgakakis08-08} to determine the sensitivity throughout the survey field. The source extraction process in {\em AE} estimates the Poisson probability that the observed counts in the detection cell arise completely from random fluctuations of the background. The three parameters that define this process are the size and shape of the detection cell and the Poisson probability threshold $P_{thresh}$ ({\em AE} value $pns$). The size and shape of the detection cell are defined by {\em AE} for each source based on the count distribution, source crowding, off-axis angle. By setting $P_{thresh}$ to $1\times10^{-2}$ for our analysis, which is the {\em AE} default value, we set the minimum number of photons required in a detection cell to be considered a source.

With the parameters of the source extraction process in hand, two additional data products are needed to proceed: an exposure map and a background map. The exposure map is the total exposure at any point in the field. We created 3 exposure maps from our existing data products for 3 separate regions of the M31 field: the northeast, bulge, and southwest. In the bulge, we merged the ACIS-I/S exposure maps using the \CIAO\ tool $\texttt{reproject\_image}$ to reproject the maps to a common tangent plane and merge them. The background maps were required to be identical in size to the exposure maps and so we generated 3 maps for each of the regions as for the exposure maps. 

\begin{figure*}
\begin{tabular}{cc}
\includegraphics[width=1.0\columnwidth]{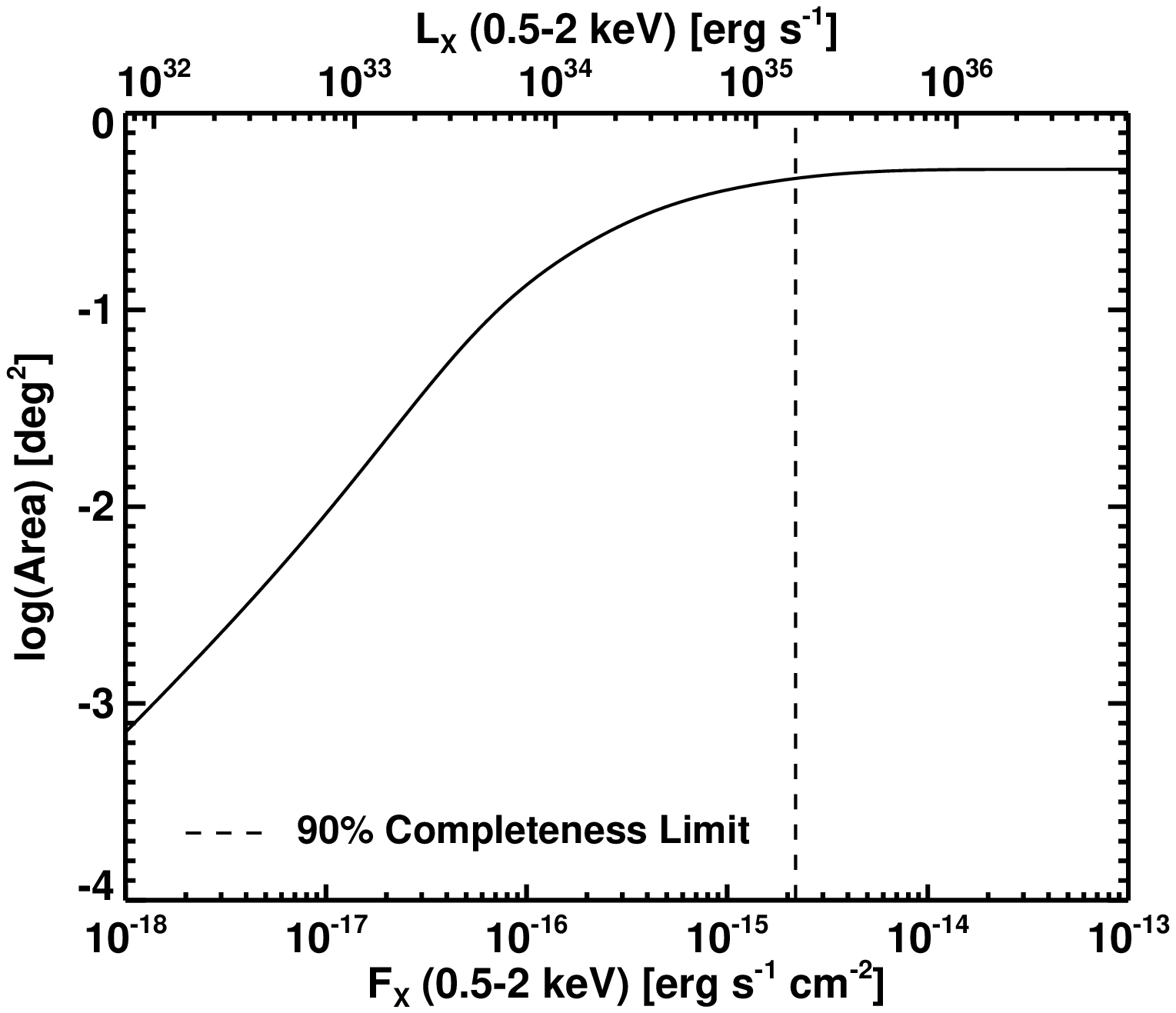}
\includegraphics[width=1.0\columnwidth]{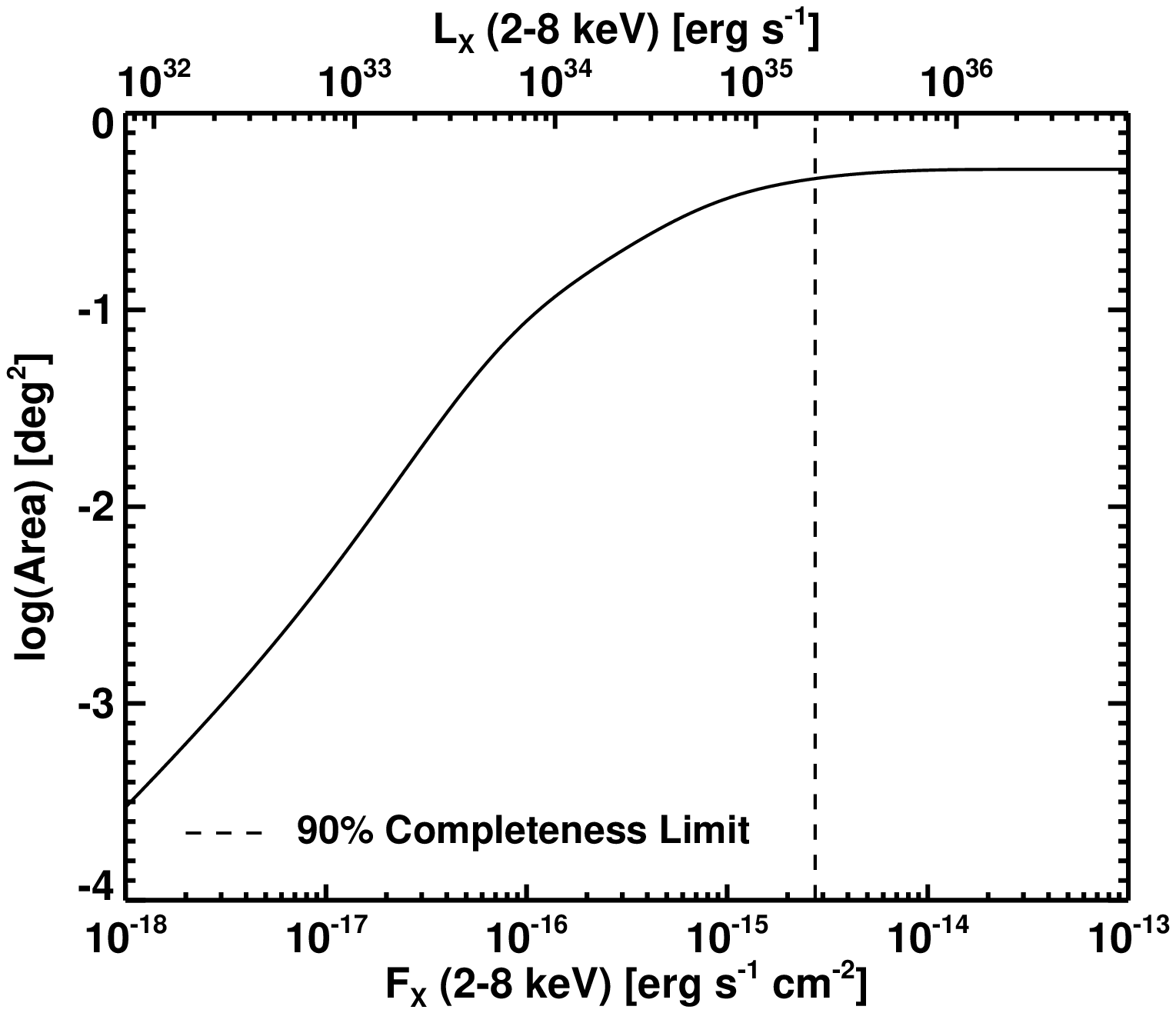}
\end{tabular}
\caption{Sensitivity curves in the soft and hard energy bands for the complete \Chandra\ survey area in M31. Sensitivity for the various overlapping observations and different fields was calculated using the method of \citet{georgakakis08-08}. Unabsorbed fluxes and luminosities are shown assuming an absorbed power-law spectrum with $\Gamma=1.7$ and $N_{H} = 6.66\times10^{20}$ cm$^{-2}$ (see Section \ref{sec:sens} for more details).} \label{fig:senscurve}	
\end{figure*}

The background map is an estimate of the source-free background across the different regions in the survey. We started by using the merged counts images in each region (created in the same manner as the merged exposure maps). We removed the counts in the vicinity of detected sources using an aperture 1.5 times larger than the 90 per cent enclosed PSF radius (obtained from {\em AE}). These pixel values were replaced by the background values calculated by {\em AE} for each source. We now had an image in units of counts where we have removed the contribution from all detected sources. However, the background map should have units of counts pixel$^{-1}$ so that every pixel in the survey field has a background value associated with it. This was required for us to estimate the sensitivity at each pixel. Such an image is created by using the \CIAO\ tool $\texttt{dmimgpm}$, which calculates a modified Poissonian mean for each pixel using a box of width 64\arcsec\ as the sampling region around that pixel. 

Using the background map values we estimated the minimum number of counts required in each pixel for a detection (above our threshold $P_{thresh}$) with the inverse survival function (Python version 2.7.8; scipy.stats.poisson module version 0.14.0). We then estimated the mean expected number of counts ($T = S + B$) that would be produced at each pixel for a range of fluxes. $B$ is the background value of the pixel and $S$ is the mean expected source contribution (equation (5) of \citealt{georgakakis08-08}). Finally, we estimated the probability that a given flux would produce the mean expected number of counts above the minimum counts required for a detection in each pixel using the survival function (python scipy.stats.poisson module). To correct our log$N$-log$S$ relations to represent the number of sources at a given flux per deg$^{-2}$ we created an area curve. This was accomplished by summing the probabilities for each pixel at a given flux to obtain the total area, which when done for the various flux values in the soft and hard energy bands gave the area curves shown in Figure \ref{fig:senscurve}.

As in \citet{stiele10-11}, we assumed an absorbed power-law spectrum with $\Gamma=1.7$ and chose $N_{H} = 6.66\times10^{20}$ cm$^{-2}$ \citep{dickey-90}, which was the weighted average for a 1 degree radius cone around the M31 nucleus. This model matches what we expect from XRBs and background AGN in M31. As \citet{tullmann04-11} point out, the model fails for extremely hard (soft) sources by overestimating (underestimating) the flux, but does not bias the log$N$-log$S$ relation to systematically higher or lower fluxes. In addition, most of the previous M31 surveys have used the same $\Gamma$ value and a similar Galactic foreground absorption, therefore making comparisons more accurate. We summarize the completeness limits for our survey in Table \ref{tab:comp}.

\begin{table}
\caption{Completeness Limits\label{tab:comp}}
\resizebox{\columnwidth}{!}{%

\begin{tabular}{@{}cccc@{}}
\hline\hline
Completeness	&	Full [$0.5-8.0$ keV]	&	Soft [$0.5-2.0$ keV]	&	Hard [$2.0-8.0$ keV]	\\
\hline
50\% &  0.66 &  0.20 &  0.33 \\
70\% &  1.37 &  0.45 &  0.69 \\
90\% &  4.05 &  1.58 &  1.98 \\
95\% &  6.73 &  2.68 &  3.29 \\
\hline
\end{tabular}
}
\begin{list}{}{}
\item Completeness limits in various energy bands calculated from our sensitivity curves. Values show unabsorbed luminosities in units of $10^{35}$ \es.
\end{list}
\end{table}

\subsection{The \logns\ Relation and X-ray Luminosity Functions} \label{sec:lognlogs}

\begin{figure*}
\begin{tabular}{cc}
\includegraphics[width=1.0\columnwidth]{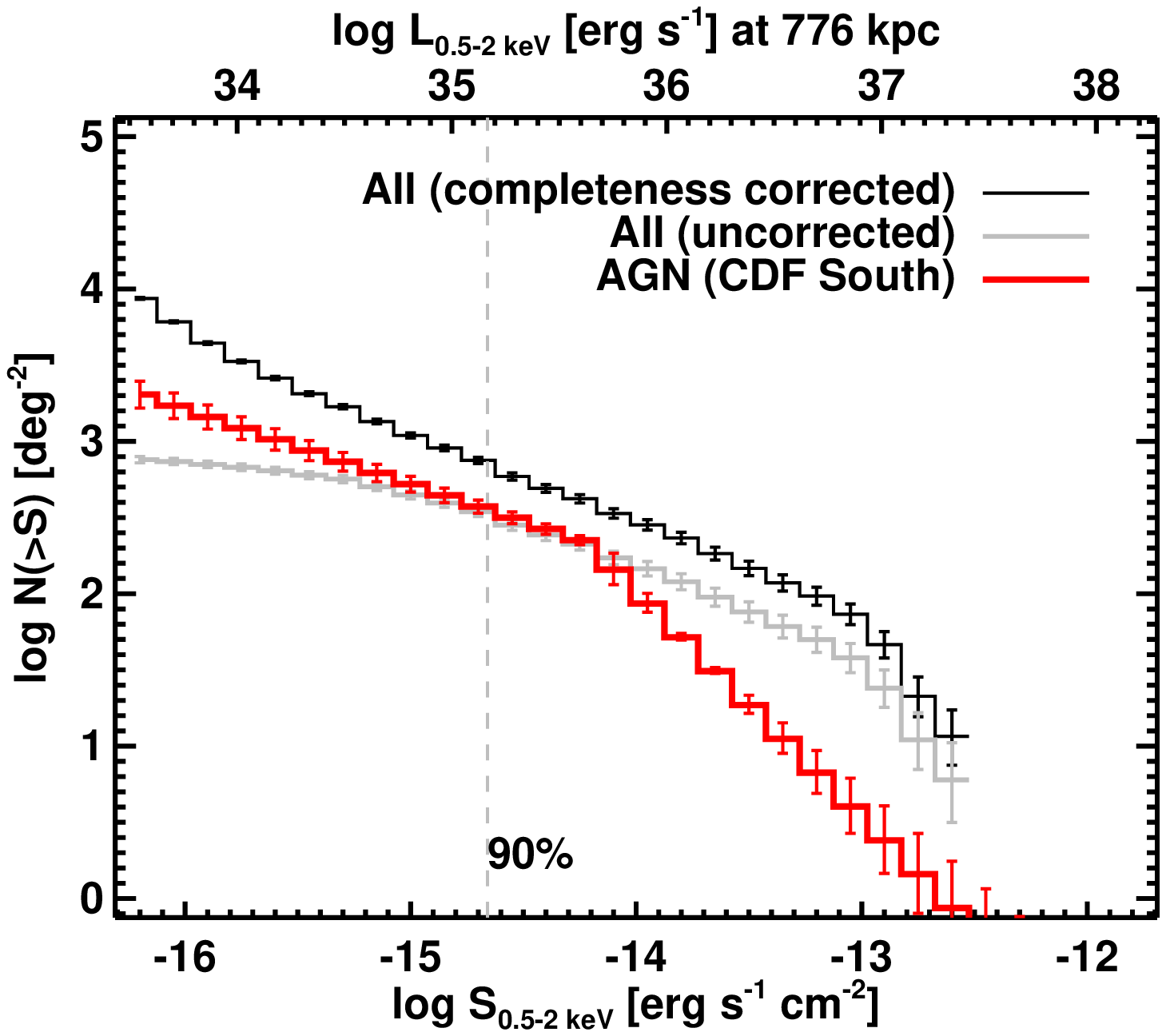}
\includegraphics[width=1.0\columnwidth]{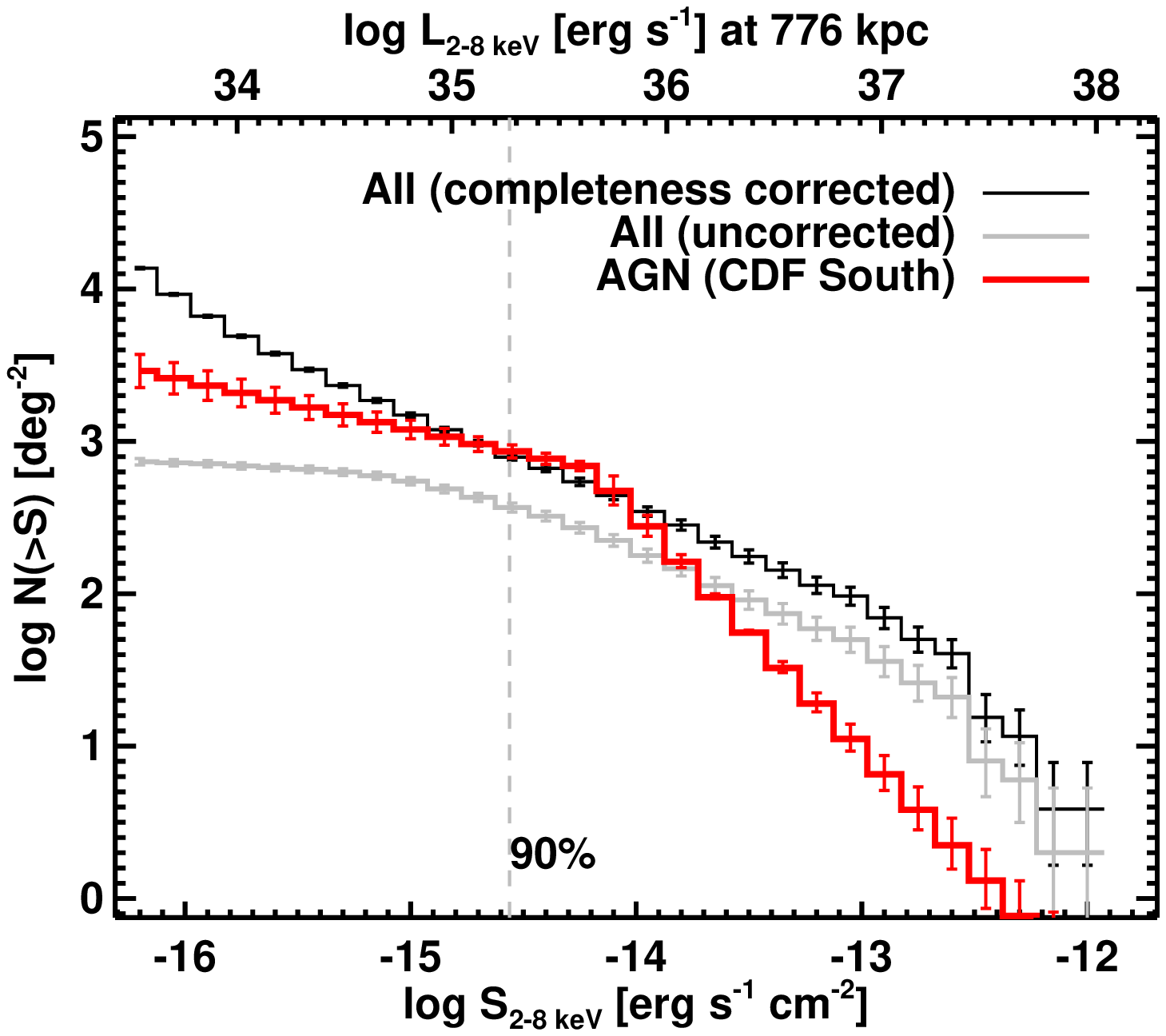}
\end{tabular}
\caption{Cumulative number counts for the soft and hard energy bands. Both uncorrected (grey) and completeness-corrected curves (black) are plotted using all X-ray point sources with calculated fluxes in the given energy band from our catalogue. All uncertainties are $1\sigma$ \citep{gehrels04-86}. The AGN number counts from the 4 Ms \Chandra\ Deep-Field South survey \citep{lehmer06-12} are shown in red with their corresponding uncertainties. The 90\% completeness limit is marked by the dashed vertical line. The red curve overtakes the black curve around the 95\% completeness limit in the hard band, indicating that the majority of sources near this flux are AGN. The flattening observed (where the XLF turns over) begins at \flatten\ \es\ and is consistent with previous results from M31 \citep{kong10-02}.}\label{fig:logns}
\end{figure*}

The \logns\ relation is calculated by determining the cumulative number of sources $N(>S)$ above a given flux $S$ (in erg s$^{-1}$ cm$^{-2}$) that are found in a survey with a total geometric area $A$:
\begin{align}
N(>S) & = \sum_{S_{i}>S}\frac{1}{A(S_{i})}
\end{align}
The value $N(>S)$ has units of sources deg$^{-2}$ and is weighted by the survey area over which a source with flux $S$ could have been detected. We constructed \logns\ distributions for the soft ($0.5-2.0$ keV) and hard ($2.0-8.0$ keV) bands using all \num\ sources in our catalogue with calculated fluxes in the given band. These XLFs are shown in Figure \ref{fig:logns}. We show both the uncorrected (grey) and completeness-corrected (black) data, which indicates the level of completeness of our survey in each band at the turnover of the grey curve, similar to the values in Table \ref{tab:comp}. We calculated $1\sigma$ uncertainties for our XLFs using Poisson statistics \citep{gehrels04-86}. We have only accounted for foreground (Galactic) extinction in computing our fluxes and therefore any internal extinction in M31 would push our curves right towards brighter fluxes. We also plot the expected contribution from AGN using the distribution of AGN from the 4 Ms \Chandra\ Deep-Field South (CDF-S) survey \citep{lehmer06-12}. The curve is shown in red with corresponding uncertainties. As expected, the hard-band XLF shows a significant contribution from expected AGN across the entire flux range. Near the 95 per cent completeness limit of $3.29\times10^{35}$ \es, the AGN contribution overtakes the corrected curve, possibly a result of cosmic variance from the CDF-S survey.

\begin{figure*}
\begin{tabular}{cc}
\includegraphics[width=1.0\columnwidth]{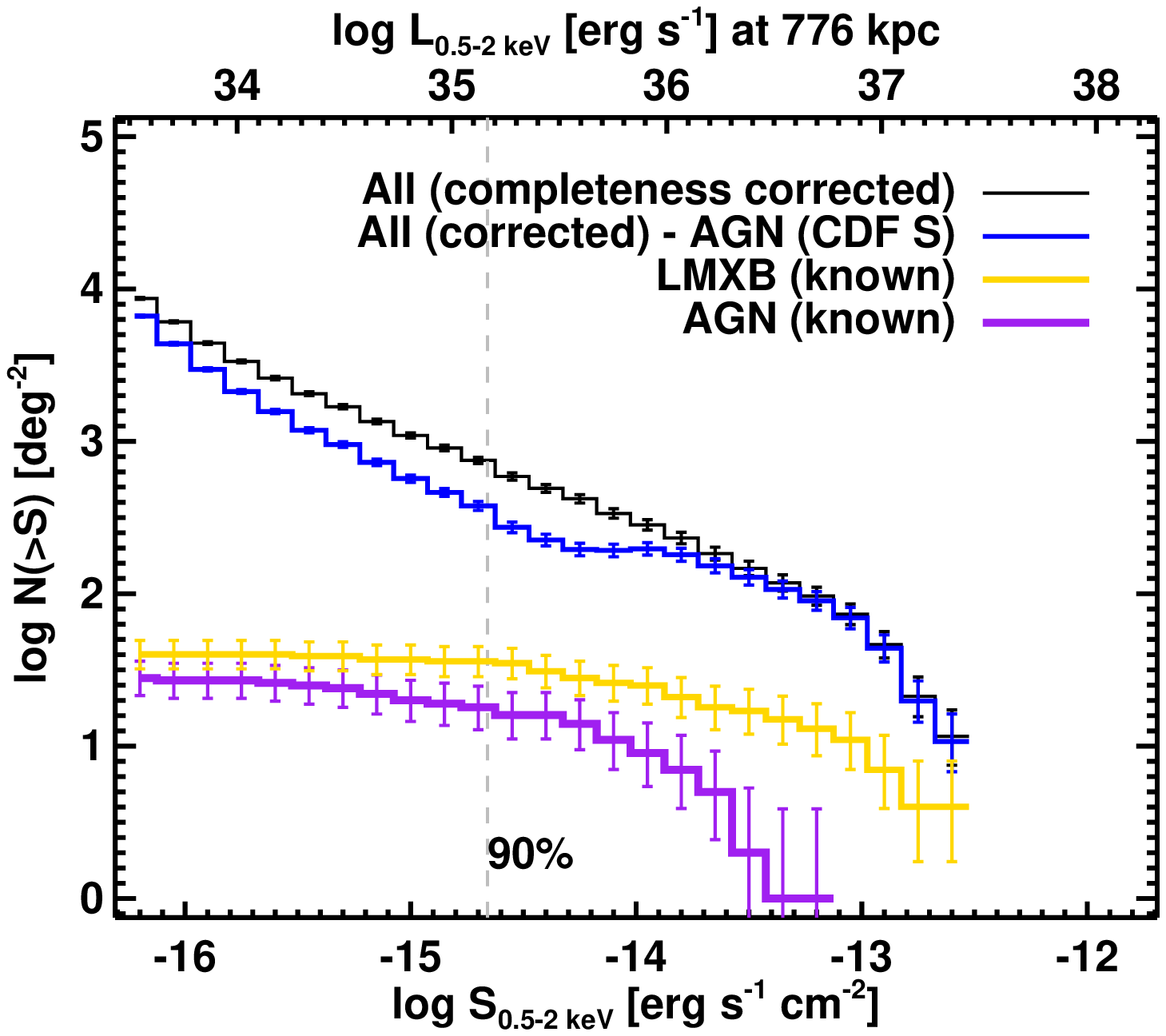}
\includegraphics[width=1.0\columnwidth]{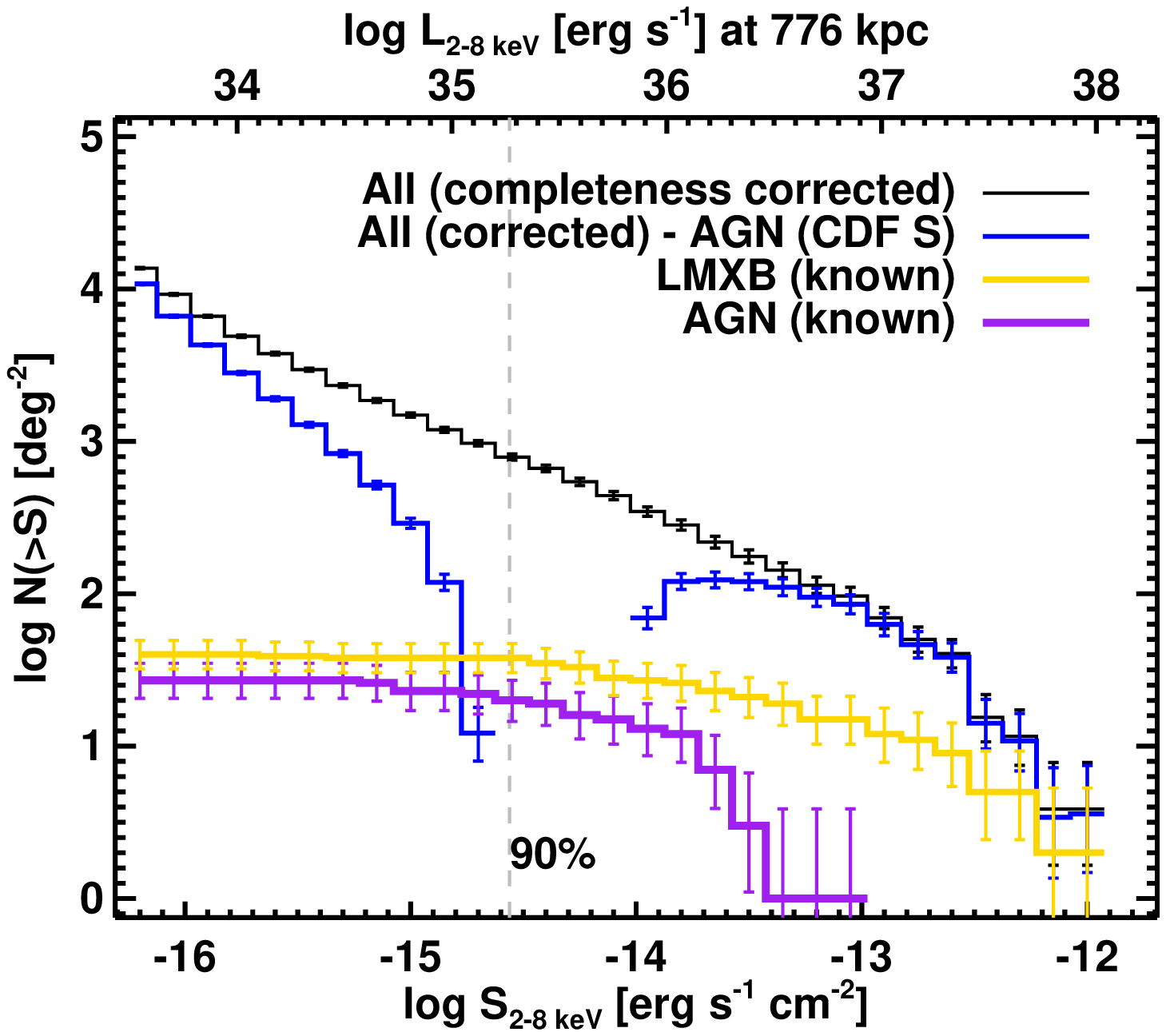}
\end{tabular}
\caption{Cumulative number counts for the soft and hard energy bands. The completeness-corrected curve (black) is plotted using all \num\ X-ray point sources from our catalogue. Known LMXBs have been matched to within 1\arcsec\ of our point sources and their XLF is shown in gold. We also used various observational catalogues of known AGN and background galaxies to match within 1\arcsec\ of our point sources, with their XLF shown in purple. The blue curve shows our completeness-corrected curve after subtracting the contribution from AGN CDF-S (red curve from Figure \ref{fig:logns}). All uncertainties are $1\sigma$ \citep{gehrels04-86}. The 90\% completeness limit is marked by the dashed vertical line. As \citet{stiele10-11} pointed out, many sources in M31 have no confirmed optical counterparts; this is reflected in the lack of known sources. The background AGN population as measured from the CDF-S accounts for all of the sources in our flux range in both the soft and hard bands. The gap in the blue curve in the hard band is likely due to the uncertainties in the CDF-S, which do not include cosmic variance, biasing the results from the CDF-S (our 95\% completeness limit is also located in this region).}\label{fig:logns2}
\end{figure*}

In Figure \ref{fig:logns2} we plot the AGN-subtracted completeness-corrected curve (blue) to represent the X-ray sources that belong to M31. We matched our detected X-ray sources in each band to a catalogue of LMXBs in M31 to within 1\arcsec. The catalogue consisted of 26 confirmed globular cluster X-ray sources and 10 confirmed field LMXBs from \citet{stiele10-11}, along with 45 confirmed LMXBs from \citet{peacock10-10}. The XLF of known LMXBs is shown in gold. We cannot produce an XLF curve for HMXBs because none have been confirmed in M31.
We also used several catalogues of background galaxies/AGN to match to our X-ray point sources within 1\arcsec. These catalogues include the 2270 background galaxies from the PHAT survey \citep{johnson04-15}, 1870 background quasars identified with the LAMOST survey \citep{huo07-10,huo06-13,huo08-15}, and radial searches of 2 degrees around the M31 centre in NED, SIMBAD, and SDSS DR12. The known AGN are represented by the purple curve in Figure \ref{fig:logns}. In Table \ref{tab:srclist} we have indicated if one of our catalogue sources was matched to an AGN or LMXB. We have also added the detailed cross-match information with the various AGN/LMXB catalogue identifications (where available) to Table \ref{tab:crossmatch}. In both energy bands, the brightest sources have a significant contribution from LMXBs, whereas the known AGN are only identified for fainter fluxes. The break in the XLFs is observed at \flatten\ \es, similar to results from previous work.
\citet{stiele10-11} found that $\sim65$ per cent of sources in their M31 catalogue had no confirmed optical counterparts, which is reflected in the lack of known LMXB/AGN sources. The gap in the AGN-subtracted completeness-corrected curve (blue) in the hard band occurs in the region of our 95\% completeness limit as well as where the CDF-S curve turns over. The background AGN population as measured from the CDF-S accounts for all of the sources in our flux range in both the soft and hard bands. However, the gap in the blue curve in the hard band is likely due to the uncertainties in the CDF-S that do not include cosmic variance, biasing the results from the CDF-S.

\begin{figure*}
\begin{tabular}{cc}
\includegraphics[width=1.0\columnwidth]{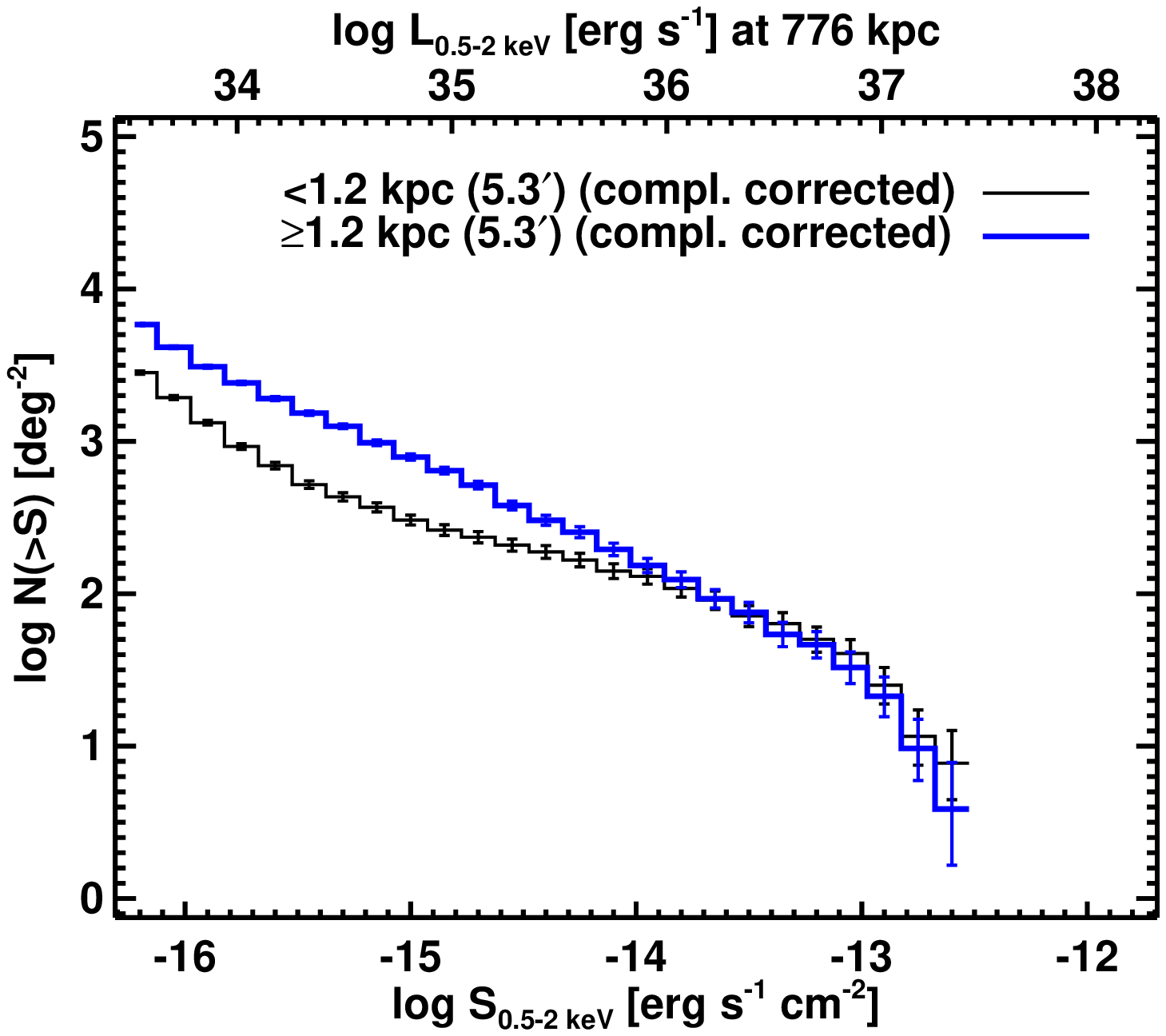}
\includegraphics[width=1.0\columnwidth]{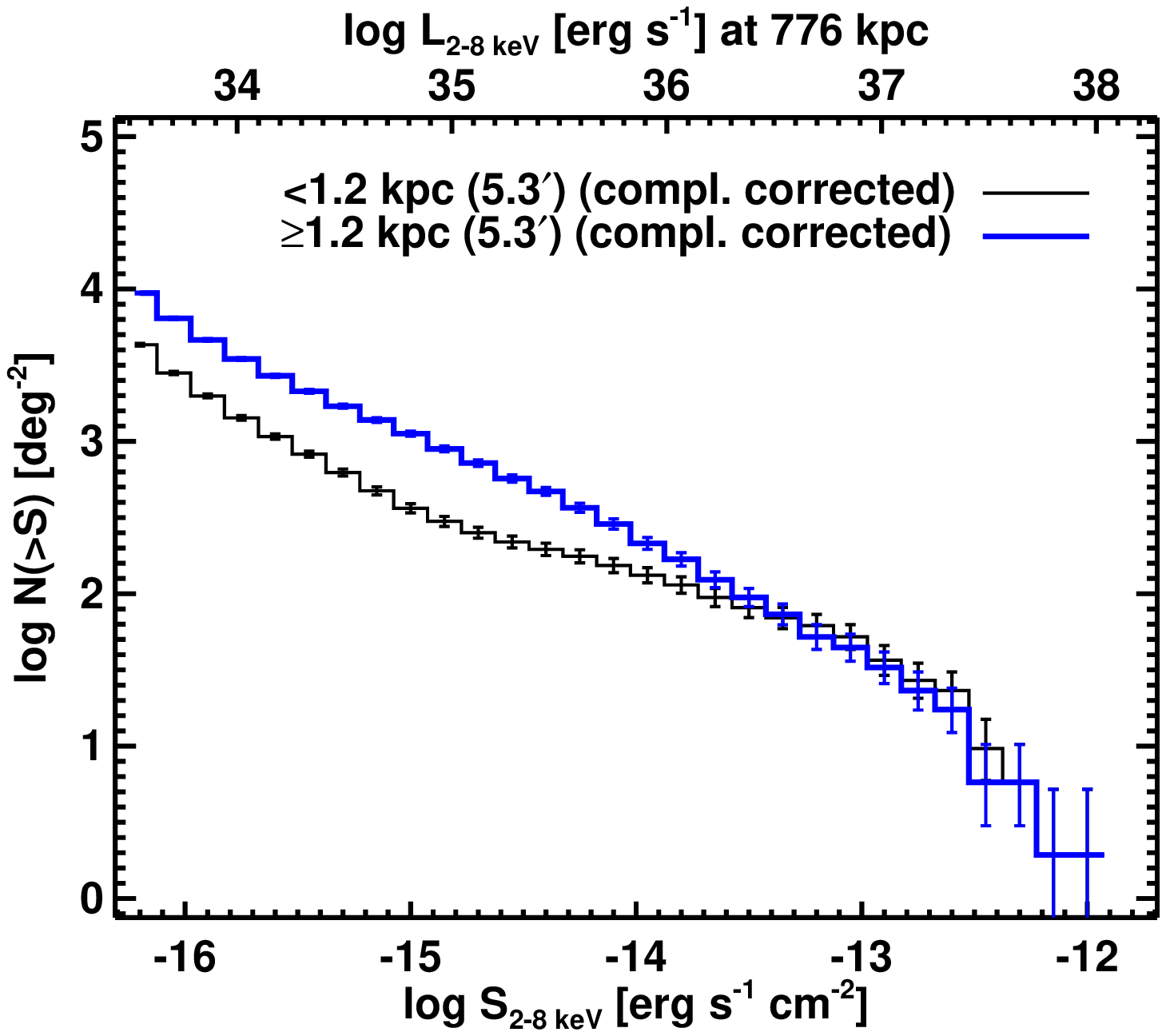}
\end{tabular}
\caption{Cumulative number counts for the soft and hard energy bands in the bulge (black) and disk$+$halo (blue) fields of M31. The completeness-corrected curves have uncertainties of $1\sigma$ \citep{gehrels04-86}. The bulge and disk$+$halo populations were separated using projected radii based on the results of \citet{courteau09-11}. In the soft band, the bulge has a larger number of brighter sources (although this may be biased due to the incompleteness of disk data), which is unexpected for an older stellar population. However, both curves are similar within uncertainties near the bright-end. In the hard band the disk$+$halo harbours the brightest sources but again the bulge has a larger cumulative number of bright sources above the break at $\approx1.3\times10^{37}$ \es. The flatter bulge XLFs in both bands are consistent with previous studies.}\label{fig:logns-bd}
\end{figure*}

In addition to our XLFs for all our sources, we also divided our sample to study the bulge and disk fields in M31. In Figure \ref{fig:logns-bd}, we show the soft and hard band XLFs for these regions. \citet{courteau09-11} found that the M31 bulge light dominates for $R_{min} \lesssim 1.2$kpc, which represents the projected minor axis radius. The disk dominates in the range 1.2 kpc $< R_{min} <$ 9 kpc, and the halo beyond that. We include the disk$+$halo curve in our XLF (represented by the $R_{min} >1 .2$ kpc curve) because there is no change in the XLF when excluding the halo sources as defined in \citet{courteau09-11}. In the soft band, the bulge has a larger number of brighter sources (this may be biased due to the incompleteness of disk data), although the number is still within the statistical uncertainties compared to the disk$+$halo. In the hard band, the disk$+$halo harbours the brightest sources but again the bulge has a larger cumulative number of bright sources above the break at $\approx1.3\times10^{37}$ \es. The flatter bulge XLFs in both bands are consistent with previous results from M31 \citep[e.g.][]{kong10-02, williams07-04}. The M31 XLF is different from other galaxies \citep[e.g.][]{colbert02-04, binder10-12} in that the older stellar population in the bulge has a flatter slope than the younger stellar population in the disk. The lack of bright sources and steeper XLF in the disk indicates that star formation in the disk is low, and thus any HMXBs would be faint (likely a reason none have yet been confirmed). 
Using \xmmn, \citet{trudolyubov05-02} analysed the bulge and two northeast fields and found the disk and bulge XLFs to have a similar slope, but using a 15\arcmin\ radius for the bulge. They did find that disk sources are all fainter than $2\times10^{37}$ \es, while brighter bulge sources existed.

XLF's have been produced for the M31 X-ray population in a number of different \Chandra\ surveys. \citet{kong10-02} found 204 sources in the bulge to a detection limit of 10$^{35}$ \es\ (no mention of a completeness limit), whereas we have about a factor of $\sim1.5$ more at the same limit in the full band (uncorrected). However, they do not correct for completeness so a direct comparison is difficult, particularly since the total areas vary. \citet{kong10-02} also found that the flattening of the XLF below $10^{36}$ \es\ in the inner bulge (2\arcmin\ by 2\arcmin) was intrinsic and not due to the incompleteness of their survey. The same flattening is seen in our XLFs albeit at a higher flux, beginning at \flatten\ \es, after correcting for completeness. The discrepancy arises from the different definition for the bulge, where ours was much larger (5.3\arcmin\ radius). \citet{kong03-03} extended their own work to three disc fields in M31, and when combining this with their data from the bulge they recovered $\sim200$ sources at a completeness limit of $10^{36}$ \es. We have 315 sources (uncorrected XLF) at this level, but it is again difficult to compare values since their XLF is not completeness-corrected. They also found that the bulge (central 17\arcmin\ square) XLF was flatter than the disk fields they surveyed. The sources in the bulge were more luminous than the outer disk fields only when removing globular cluster X-ray sources from disk fields. 
\citet{williams07-04} detected 85 sources above $4\times10^{36}$ \es, while we detected twice as many above the same limit. They also found that the disk XLF was steeper than the bulge with a power law slope comparable to typical elliptical galaxies. We detected a larger number of sources due to variations in sensitivity, coverage (M31 X-ray source density varies with radius), and source extraction parameters. While these factors all contributed to some degree, the transient nature of X-ray sources, specifically binary systems, is the primary reason for such a discrepancy. This was demonstrated by the matching results from Section \ref{sec:xmm}, where \leftoverxmm\ and \leftover\ \xmmn\ and \Chandra\ sources respectively were unmatched in the same field of view. Sensitivity is a factor, however, since \xmmn\ and \Chandra\ surveys did not have equal exposure throughout, and \Chandra\ is generally more sensitive.

Our XLFs are the deepest in terms of luminosity produced for any large galaxy based on our detection limits. The AGN contribution from the CDF South dominates the hard band XLF as expected, whereas a larger proportion of sources in the soft band are representative of M31 X-ray sources (e.g.\ LMXBs, supernova remnants). 
Based on the distribution of AGN from the CDF-S, the M31 contribution observed from various catalogues is incomplete below $5\times10^{-13}$ \esc\ in the hard band. The PHAT survey has identified many AGN by-eye using crowd-sourcing and criteria that led to a minimal number of misclassifications (Section 3 of \citealt{johnson04-15}). However, the field of view of these AGN did not overlap well with the \Chandra\ data in this catalogue. As in the Milky Way, M31 has only a handful of very bright X-ray sources (mostly in globular clusters), and none above $\approx10^{38}$ \es. Due to M31's low star formation rate, very few bright HMXBs would be expected, and to date only $\sim30$ candidates have been identified.

\section{Summary}	\label{sec:summary}

We have used 133 publicly available \Chandra\ ACIS-I/S observations totalling $\sim1$ Ms to create the deepest X-ray point source catalogue of M31. We detected \num\ X-ray sources within our field of view (\area\ deg$^{2}$) to a limiting unabsorbed $0.5-8.0$ keV luminosity of $\sim10^{34}$ \es. Our 90\% ($0.3-8.0$ keV) completeness limit is $4\times10^{35}$ \es. We detected \blg, \nes, and \sw\ sources in the bulge, northeast, and southwest fields of M31 respectively. In the bulge fields, X-ray fluxes are closer to average values because they are calculated from many observations over a long period of time. Similarly, our catalogue is more complete in the bulge fields since monitoring allows more transient sources to be detected. Cross-correlating our catalogue with a previous \xmmn\ catalogue of 1948 X-ray sources, with only 979 within the field of view of our survey area, we found \xcmatch\ (\matchperall\%) of our \Chandra\ sources (\xcmatchuniq\ or \matchper\% were unique sources) matched to within 5\arcsec\ of \xcmatchuniq\ \xmmn\ sources with a median offset of \medsepx\arcsec. Similarly, we matched our catalogue to a master list of \allprevc\ previously published \Chandra\ sources in M31 and found \catc\ of our sources within 1\arcsec. Collating the matching results from all catalogues we found \absnew\ new sources in our catalogue. We also created XLFs in the soft and hard bands that are the deepest for any large galaxy based on our detection limits. Using published catalogues of AGN and LMXBs we determined the contribution to the XLF from these populations. The observationally identified AGN in M31 are incomplete below $\sim10^{-13}$ \esc\ (hard band) based on data from the \Chandra\ Deep Field South. The completeness-corrected XLFs show a break at \flatten\ \es, which is consistent with previous work in M31. We found that the bulge XLFs are flatter compared to the disk, consistent with other studies. This indicates a lack of bright high-mass X-ray binaries in the disk due to a low star formation rate and an aging population of low-mass X-ray binaries in the bulge. This catalogue is more robust and complete than the latest \Chandra\ source catalogue release due to the stringent processing and requirements we have placed on source detection. In addition, we have published a much more detailed set of source characteristics using {\em ACIS Extract}.

Impending \Chandra\ and \nustar\ X-ray surveys in M31 will cover new regions of the galaxy (e.g. PHAT field) that have only been observed by \xmmn. This will result in high spatial resolution $0.3-30$ keV data that will be crucial for classifying and characterising the X-ray source population.

\section*{acknowledgements}
We thank the referee for valuable comments that improved the manuscript. We thank Ben Williams for a CFHT image of the PHAT region of M31 used for astrometry and for helpful comments on the manuscript. We thank Cliff Johnson for the PHAT footprint. We thank Pat Broos and Leisa Townsley for all their expertise and advice on optimising {\em AE}. We also thank Antonis Georgakakis for providing us with his code for calculating a sensitivity map. Support for this work was provided by Discovery Grants from the Natural Sciences and Engineering Research Council of Canada and by Ontario Early Researcher Awards. NV acknowledges support from Ontario Graduate Scholarships. We have also used the Canadian Advanced Network for Astronomical Research (CANFAR; \citealt{gaudet07-11}) and thank S\'{e}bastien Fabbro for all his help. This work was made possible by the facilities of the Shared Hierarchical Academic Research Computing Network (SHARCNET:www.sharcnet.ca) and Compute/Calcul Canada. We thank Mark Hahn for all his efforts to appease our processing needs. This research has made use of the NASA/IPAC Extragalactic Database (NED), which is operated by the Jet Propulsion Laboratory, California Institute of Technology, under contract with the National Aeronautics and Space Administration. This publication makes use of data products from the Two Micron All Sky Survey, which is a joint project of the University of Massachusetts and the Infrared Processing and Analysis Center/California Institute of Technology, funded by the National Aeronautics and Space Administration and the National Science Foundation. This work also made use of MARX \citep{davis09-12}. We acknowledge the following archives: the Hubble Legacy Archive (\url{hla.stsci.edu}), Chandra Data Archive (\url{cda.harvard.edu/chaser}), and 2MASS (\url{ipac.caltech.edu/2mass}). \\ 
\indent \emph{Facilities:} HST (ACS, WFC3), CXO (ACIS)

\bibliographystyle{mnras}

\bsp

\end{document}